\documentclass[11pt]{article}

\pdfoutput=1

\usepackage{jheppub} 
\usepackage{amsmath,amssymb,epsfig,amsfonts}

\usepackage{epsf}
\usepackage{makeidx}

    \usepackage{slashed}
    \usepackage{verbatim} %for comments
    \usepackage{float}
    \restylefloat{table}
    \usepackage{pdflscape}
    \usepackage{tikz}
    \usepackage{pifont}
\usepackage{array}
    \usepackage{youngtab}
    \usepackage{multirow}
    \usepackage{xcolor}
    \usepackage{ulem}
    \usepackage{cleveref}
    \usepackage{tensor}
    \usepackage{todonotes}
    \usepackage{mathrsfs}
    \usepackage{changepage}
    \usepackage{anyfontsize}
   \usepackage{caption}
\usepackage{subcaption}

\makeatletter

\newcommand{\be}{\begin{equation}}
\newcommand{\ee}{\end{equation}}
\newcommand{\ba}{\begin{aligned}}
\newcommand{\ea}{\end{aligned}}

\newcommand{\lp}{\ell_{\text{p}}}
\newcommand{\dd}{\mathrm{d}}

\newcommand{\me}{\mathrm{e}}
\newcommand{\ii}{\mathrm{i}}
\newcommand{\vol}{\mathrm{vol}}

\newcommand{\ls}{\ell_s}
\newcommand{\dz}{\gamma}
\newcommand{\lsc }{\Xi}

\makeatother

\begin{document}

% format
\baselineskip=18pt  % a la harvmac
\numberwithin{equation}{section}  % make eq labels (sec.num)
\allowdisplaybreaks  % allow page breaks in displayed eqs

%%%%%%%%%%%%%%%%%%%%%%%%%%%%%%%%%%%%%%%%%%%
%%%        TITLE BEGINS HERE
%%%%%%%%%%%%%%%%%%%%%%%%%%%%%%%%%%%%%%%%%%%

%% ========== title (note version) begins here ==========
%
%\vspace*{-1cm}
%\begin{center}
% {\Large\bf Title of the Document}
%\end{center}
%\vspace*{-.5cm}
%
%% ========== title (note version) ends here ==========

%% ========== title (paper version, a la harvmac) begins here ==========

\thispagestyle{empty}

% title, authors, affiliation
\vspace*{1cm} 
\begin{center}

{\LARGE  Holographic duals of M5-branes on an irregularly punctured sphere } 
 \vspace*{1.8cm}

{\fontsize{12.3}{17}\selectfont Christopher Couzens$^{}$\footnote{cacouzens@khu.ac.kr}, Hyojoong Kim$^{}$\footnote{h.kim@khu.ac.kr}, 
 Nakwoo Kim$^{}$\footnote{nkim@khu.ac.kr}, Yein Lee$^{}$\footnote{lyi126@khu.ac.kr}
}

 \vspace*{.5cm} 
{${}^{}$ Department of Physics and Research Institute of Basic Science,\\
  Kyung Hee University, Seoul 02447, Republic of Korea\\}
  
 {\tt {}}

\vspace*{0.8cm}
\end{center}

 \renewcommand{\thefootnote}{\arabic{footnote}}
 
\begin{center} {\bf Abstract } \end{center}
\noindent
We provide explicit holographic duals of M5-branes wrapped on a sphere with one irregular puncture and one regular puncture of arbitrary type. The solutions generalise the solutions corresponding to M5-branes wrapped on a disc recently constructed by Bah--Bonetti--Minasian--Nardoni by allowing for a general choice of regular puncture. We show that the central charges, flavour central charges and conformal dimensions of BPS operators match with a class of Argyres--Douglas theory.

\noindent 

%\vspace*{.5cm}

% abstract
%\noindent

\newpage
%%%%%%%%%%%%%%%%%%%%%%%%%%%%%%%%%%%%%%%%%%%
%%%           TITLE ENDS HERE
%%%%%%%%%%%%%%%%%%%%%%%%%%%%%%%%%%%%%%%%%%%

\tableofcontents
\printindex
\setcounter{footnote}{0}

%%%%%%%%%%%%%%%%%%%%%%%%%%%%%%%%%%%%%%%%%%%
%%%        MAIN TEXT BEGINS HERE
%%%%%%%%%%%%%%%%%%%%%%%%%%%%%%%%%%%%%%%%%%%
\newpage

\section{Introduction}
Compactifying SCFTs on compact manifolds has been a fruitful avenue for constructing new SCFTs. Given a parent SCFT with a holographic dual, it is natural to consider the holographic dual of the compactified theory. The earliest examples of such dual pairs were constructed in \cite{Maldacena:2000mw}, and studied the compactification of both 4d $\mathcal{N}=4$ super-Yang--Mills and the 6d $\mathcal{N}=(2,0)$ theory on a Riemann surface with genus $g>1$. Since then this avenue of research has been extended in multiple directions. In this work we will primarily be interested in 4d $\mathcal{N}=2$ theories that can be obtained from M-theory. 
Theories of class $\mathcal{S}$ are 4d $\mathcal{N}=2$ SCFTs which arise from wrapping the 6d $\mathcal{N}=(2,0)$ theory living on M5-branes on a punctured Riemann surface with gravity duals classified in \cite{Gaiotto:2009gz,Gaiotto:2009we}\footnote{The type IIA picture is given in \cite{Reid-Edwards:2010vpm,Aharony:2012tz}.}. Extensions to 4d $\mathcal{N}=1$ theories from wrapped M5-branes have been studied in \cite{Gauntlett:2004zh,Agarwal:2014rua,Bah:2013aha,Bah:2012dg}.

The prototypical setup for studying branes wrapped on various compact cycles is to consider embedding the cycle in some larger geometry, for example a Calabi--Yau or G2 manifold and taking the metric on the compact space to be the one with constant curvature. Given that we typically view these types of solutions as the IR fixed point of an RG flow interpolating between dimensions, or rather the near-horizon of some black object, one can motivate this ansatz, at least for Riemann surfaces, from known uniformization theorems \cite{Anderson:2011cz,Bobev:2020jlb}. Under some assumptions, they state that one can take the UV solution to have an arbitrary metric on the Riemann surface, and along the RG flow this is washed out leaving just the constant curvature metric at the IR fixed point. Key assumptions for these theories are that supersymmetry is realised by a topological twist and that the metrics are smooth.

Recently, solutions with non-constant curvature metrics have been studied in string and M-theory, thus evading the uniformization theorems. These solutions are known as spindles \cite{Ferrero:2020laf,Hosseini:2021fge,Boido:2021szx,Faedo:2021kur,Cassani:2021dwa,Ferrero:2021ovq,Couzens:2021rlk,Faedo:2021nub,Ferrero:2021etw,Couzens:2021cpk,Giri:2021xta} and discs \cite{Bah:2021mzw,Bah:2021hei,Couzens:2021tnv,Suh:2021ifj,Suh:2021hef,Suh:2021aik,Couzens:2021rlk,Karndumri:2022wpu} and can be viewed as the horizon of accelerating black objects.\footnote{For discussion on accelerating black holes not embedded in string or M-theory see for example \cite{Anabalon:2018qfv,Anabalon:2018ydc,Gregory:2017ogk} and references within.} Spindles are the orbifold $\mathbb{WCP}^{1}_{n_-,n_+}$, i.e. a two-sphere with conical deficits at both poles. Discs on the other hand have the topology of (unsurprisingly) a disc, with an orbifold singularity at the centre and a boundary on which the metric is locally a cylinder but singular. The two types of solutions are intimately related and have been shown to be different global completions of the same local solutions \cite{Couzens:2021rlk,Couzens:2021tnv}. Apart from having metrics of non-constant curvature one of the interesting features of these geometries is the method in which supersymmetry is preserved. The mechanism is not the usual topological twist but instead requires a mixing between the parent R-symmetry and the isometry direction of the compactification space. For spindles there are two different types of twist, the anti-twist or the topological topological twist \cite{Ferrero:2021etw}, while for discs a different mechanism involving a holonomy for a gauge field on the boundary of the disc allows for the preservation of supersymmetry. See also \cite{Cheung:2022ilc,Couzens:2022lvg} which consider 4d orbifolds with non-constant curvature. 

In this work we will be interested in extensions of disc solutions, specifically for M5-branes wrapped on a disc. In \cite{Bah:2021mzw,Bah:2021hei} (BBMN) the first disc solution was presented and a dual field theory was proposed. The dual field theory is a 4d $\mathcal{N}=2$ SCFT of Argyres--Douglas type and constitutes the first holographic dual for such a theory. The dual field theory was shown to be $(I_{\hat{N},\hat{k}},Y_{l})$, which is the theory on a stack of M5-branes of A-type wrapped on a twice punctured sphere, with one irregular puncture of type I at one pole and a regular puncture of a particular type at the other pole, we will review this nomenclature in section \ref{sec:fieldtheory}. 
The goal of the present paper is to extend to more general Argyres--Douglas theories. In particular we will provide holographic duals for the SCFTs with arbitrary regular puncture and fixed type I irregular puncture. We provide evidence for the proposal by matching various observables on the two sides. 

The outline of the paper is as follows. In section \ref{sec:single} we begin by reviewing the disc solution in \cite{Bah:2021mzw,Bah:2021hei}. Our presentation uses a different set of coordinates which are better adapted for later. We embed the solution into the classification of  $\mathcal{N}=2$ AdS$_5$ solutions in M-theory in \cite{Gaiotto:2009gz,Lin:2004nb} studying the equivalent electrostatic problem. The electrostatic problem has some new and novel features, in particular unusual boundary conditions for the potential. In section \ref{sec:multi} we generalise the disc solution building on the original solution. We show that the new solutions are well-defined supergravity solutions, performing a regularity analysis and flux quantisation. We proceed to compute various observables of the gravity solution with which to match to the field theory analysis in section \ref{sec:fieldtheory}. We provide a dictionary between the field theory and gravity solution with which we can compare the observables on the two sides and find perfect agreement in the holographic limit. We relegate some material to two appendices. The first shows how to obtain AdS$_7$ and the Maldacena--Nunez solutions from our starting point, whilst the second performs an independent check of the observable computations by using anomaly inflow.

%%~~~~~~~~~~~~~~~~~~~~~~~~~~~~~~~~~~~~~~~~~~~~~~~~~~~~~~~~~~~~~~~~~~~~

%%%%%%%%%%%%%%%%%%%%%%%%%%%%%%%%%%%%%%%%%%%
%%%        BBMN Review
%%%%%%%%%%%%%%%%%%%%%%%%%%%%%%%%%%%%%%%%%%%

\section{M5-branes on a disc}\label{sec:single}

In \cite{Bah:2021hei} an AdS$_5\times\Sigma$ solution in 7d gauged supergravity was found where $\Sigma$ is topologically a disc. The boundary of the disc corresponds to a singularity of the overall metric, while the centre of the disc has a conical singularity. Upon uplifting the solution to 11d supergravity on an $S^4$ one obtains a $1/2$ BPS AdS$_5$ solution, and one can give a physical interpretation of the singularities; the boundary of the disc arises due to a stack of smeared M5-branes whereas the conical deficit is due to the presence of a monopole. In \cite{Bah:2021hei} they conjectured that the solution was holographically dual to an Argyres--Douglas theory with one regular and one irregular puncture. In the following we will review the solution, albeit from a different parametrisation, before embedding the solution into the classification of $\mathcal{N}=2$ AdS$_5$ solutions of 11d supergravity constructed in \cite{Gaiotto:2009gz,Lin:2004nb}. We use the reformulation of the solution in terms of an electrostatics problem for a single potential, determining it for the disc solution before studying its properties and flux quantisation within the electrostatics reformulation.

\subsection{AdS$_5\times \Sigma$ solutions of 7d gauged supergravity}

In this section we will study the AdS$_5\times \Sigma$ solution of 7d U$(1)^2$ gauged supergravity originally found in \cite{Bah:2021hei} using the conventions there for the gauged supergravity theory. As pointed out in \cite{Couzens:2021tnv,Couzens:2021rlk} one can obtain disc solutions as different global completion of the same local solutions from which one may construct spindle solutions. In the following we will use the parametrisation of the spindle solutions in \cite{Ferrero:2021wvk} specialised to the disc solution.\footnote{This solution specialised to preserve $\mathcal{N}=2$ may also be found in \cite{Lin:2004nb}. In fact in this reference a more general 7d solution is presented up to solving a non-trivial ODE. This generalises the solution here by including an additional scalar field. In the uplifted theory this breaks a U$(1)$ symmetry, which will be essential for our generalisation and therefore we stick with this less general solution.} The solution is
\begin{align}
\dd s^2_7&=\Big(w P(w)\Big)^{1/5}\bigg[4 \dd s^2(\text{AdS}_5)+\frac{w}{f(w)}\dd w^2+\frac{f(w)}{P(w)}\dd z^2\bigg]\, ,\nonumber\\
A_{i}&=-\frac{s_{i}}{h_i(w)}\dd z\, ,\qquad X_{i}(w)=\frac{\Big(w P(w)\Big)^{2/5}}{h_i(w)}\, ,\label{eq:7dM5sol}
\end{align}
where $\dd s^2(\text{AdS}_5)$ is the unit radius metric on AdS$_5$, satisfying $R_{\mu\nu}=-4 g_{\mu\nu}$ and the functions take the form
\begin{align}
h_{i}(w)&=w^2-s_i\, ,\qquad 
P(w)=h_1(w)h_2(w)\, ,\qquad 
f(w)=P(w)-w^3\, .
\end{align}
The solution depends on two real constants, $s_i$. In order to obtain a disc the functions $f(w)$ and $P(w)$ must have a common root. Given the form of the polynomials, it is clear that this must necessarily be at $w=0$ and therefore in order to obtain a disc we set, without loss of generality, $s_2=0$. In fact this limit leads to an enhancement of supersymmetry, with the solution preserving $\mathcal{N}=2$ supersymmetry, rather than $\mathcal{N}=1$. We will show this later by explicitly embedding the solution into the $\mathcal{N}=2$ classification of AdS$_5$ solutions of 11d supergravity in \cite{Gaiotto:2009gz,Lin:2004nb}. This local metric also admits other interesting limits. One may recover both pure AdS$_7$  and the Maldacena--Nunez solution \cite{Maldacena:2000mw} by taking different limits of this local metric as we show in appendix \ref{app:limits}.\\

\noindent {\bf Regularity in 7d}

\noindent Let us begin by considering the regime of the $s_1$ which gives rise to a well-defined metric. We require $f(w)$ to admit two roots, one at $0$ and the second positive. The latter condition is necessary for the metric to have the correct signature, since there are terms proportional to $w$ appearing in the metric. The roots of $f(w)$ are
\be
w_0=0\, ,\quad (\text{twice})\, ,\qquad w_{\pm}=\frac{1}{2}\Big(1\pm\sqrt{1+4 s_1}\,\Big)\, .
\ee
In order for there to be a real positive root and for the scalars to be positive we require 
\be
-\frac{1}{4}\leq s_1\leq0\, .
\ee
This leads to two positive roots and we take the domain of the line interval parametrised by $w$ to be
\be
w\in [0,w_-]\, .
\ee
We must now check how the metric degenerates at the end-points. 

First consider the end-point at $w=w_-$. Since $P(w_-)\neq0$ we need only consider the metric on $\Sigma$ at this end-point.
Expanding the metric on $\Sigma$ around $w=w_-$ we find
\begin{align}
\dd s^2(\Sigma)&\simeq\frac{w_-}{|f'(w_-)|(w_--w)}\dd w^2 + \frac{|f'(w_-)|}{P(w_-)}(w_--w)\dd z^2\nonumber\\
&=\frac{4 w_-}{|f'(w_-)|}\bigg[ \dd r^2 + \frac{|f'(w_-)|^2}{4 w_-^4} r^2 \dd z^2\bigg]\, ,
\end{align}
where $w_--w=r^2$. Fixing the period of $z$ to be
\be
\frac{\Delta z}{2\pi}=\frac{2 w_-^2}{l |f'(w_-)|}=\frac{2}{l (w_+-w_-)}\, ,\qquad l\in \mathbb{Z}^+
\ee
the space is the orbifold $\mathbb{R}^2/\mathbb{Z}_l$.

Let us now consider the end-point at $w=0$. Expanding the 7d metric around $w=0$ we find
\vspace{5mm}
 \begin{align}
\dd s^2_7&\simeq w^{3/5}w_-^{1/5}\bigg[4 \dd s^2(\text{AdS}_5)+\dd z^2 + \frac{1}{w w_+w_-}\dd w^2\bigg]\nonumber\\
&=r^{6/5}w_-^{1/5}~\bigg[ 4 \dd s^2(\text{AdS}_5)+\dd z^2+\frac{4}{w_+w_-}\dd r^2\bigg]\, ,
\end{align}
where we performed the change of coordinate $w=r^2$. Clearly this is singular, as one can verify by computing the Ricci scalar, or any other curvature invariant. In addition, the scalars also have a singular behaviour,
\be
X_1\sim \frac{r^{12/5}}{|s_1|^{3/5}}\, ,\qquad X_2\sim \frac{|s_1|^{2/5}}{r^{8/5}}\, .
\ee
One should contrast this singular behaviour with the singular behaviour of the 4d and 5d solutions for M2-branes \cite{Couzens:2021rlk,Suh:2021hef} and D3-branes \cite{Suh:2021ifj,Couzens:2021tnv} on discs. One notes that in addition to the singular metric only a single scalar diverges in each of these cases with the other scalars tending to zero. These scalars describe the stretching and squashing of the sphere in the uplifted theory, when written in embedding coordinates adapted to the U$(1)^n$ symmetry. In the M2-branes and D3-branes cases the sphere diverges along one direction and shrinks in the remaining directions. In contrast, here we have three scalars parametrising the squashing; two of which diverge and only one vanishes.  We will see later, using the uplifted solution, that the behaviour of M5-branes on a disc is somewhat different to that of the M2-brane and D3-brane cases. \\

\noindent {\bf Magnetic charge and holonomy}

\noindent The metric is supported by a single magneticaly charged gauge field. The magnetic charge is defined to be
\be
Q_i=\frac{1}{2\pi}\int_{\Sigma}F_{i}\, ,
\ee
which for the solution at hand is  
\be
Q\equiv Q_1=-w_-\frac{\Delta z}{2\pi}\, ,\qquad Q_2=0\, .
\ee
Note that due to the orbifold it is not necessary that $Q$ is integer but rather the weaker condition $l\, Q\in \mathbb{Z}$. As such let us define
\be
l Q=-  p \, ,\qquad p\in \mathbb{Z}^{+}\, .\label{eq:defp}
\ee
Since the disc also has a boundary one can define the holonomy of the gauge field on the boundary. One should choose a gauge for the gauge field so that it is globally well-defined on the disc. Since the circle shrinks at the centre of the disc we must require that the gauge field vanishes there. This uniquely fixes the gauge and the globally well-defined gauge field is
\be
A_1=-\bigg(\frac{s_1}{w^2-s_1}-\frac{s_1}{w_{-}^2-s_1}\bigg)\dd z\, .
\ee
The holonomy of the gauge field along the boundary is then 
\be
\text{hol}_{\partial \Sigma}(A_{1})=\frac{1}{2\pi}\oint_{\partial \Sigma}A_1=w_-\frac{\Delta z}{2\pi}=-Q\, ,
\ee
which gives minus the total magnetic charge threading through the disc. 
\\

\noindent {\bf Euler Characteristic}

\noindent
The final observable that we can compute is the Euler characteristic of the disc. 
We have
\begin{align}
\chi(\Sigma)&=\frac{1}{4\pi}\int_{\Sigma}R\dd \vol(\Sigma)+\frac{1}{2\pi}\int_{\partial \Sigma}\kappa \dd\vol(\partial \Sigma)\nonumber\\
&=\frac{\Delta z}{4\pi}\frac{w^{3/2}(3 f(w)-w f'(w))}{P(w)^{3/2}}\bigg|_{w=0}^{w=w_-}\nonumber\\
&=\frac{1}{l}\, ,
\end{align}
where in the going to the second line we have used that the geodesic curvature of the boundary of the disc is 0, it is locally a cylinder there, and hence the boundary contribution vanishes.

 Using that $w_++w_-=1$ we find the relation
\be
\chi(\Sigma)-Q=\frac{\Delta z}{4\pi}\, .
\ee
This relation is a generic feature of disc geometries, with equivalent expressions for M2-branes, D3-branes and D4-D8-brane compactifications \cite{Couzens:2021rlk,Couzens:2021tnv,Suh:2021aik,Suh:2021hef,Suh:2021ifj}. 
 
As an aside one may express everything in terms of the orbifold weight $l$ and the integer magnetic charge $p$ defined in \eqref{eq:defp}. The roots in terms of these integer parameters are
\be
w_-= \frac{p}{2(p+1)}\, ,\qquad w_+=1-w_-=\frac{p+2}{2(p+1)}\, ,\label{eq:rootsp}
\ee
and the period satisfies
\be
\dz \equiv\frac{\Delta z}{2\pi}=\frac{2(p+1)}{l}\, .\label{eq:deltaz}
\ee
It is useful for later to introduce the $2\pi$-periodic coordinate $\hat{z}$ as
\be
 \hat{ z}=\frac{z}{\dz}\, .
\ee

\vspace{4mm}

\noindent {\bf Dictionary to compare with BBMN}

\noindent
To translate between the parametrisation given above and the solution appearing in \cite{Bah:2021hei} one should perform the following identifications: 
\begin{align}
w&=\frac{B \tilde{w}}{2\sqrt{1-\tilde{w}^2}}\, ,\qquad \frac{\Delta z}{4\pi}=\mathcal{C}\, ,\qquad 4 s_1=-B^2\, ,\qquad m=1\, ,\qquad l=l_{BBMN}\, .
\end{align}
This puts the metric into the form
\be
m^2\dd s_7^{2}(BBMN)=\frac{ B \tilde{w}^{3/5}}{2\sqrt{1-\tilde{w}^2}}\bigg[4\dd s^2(\text{AdS}_5)+\frac{2}{ \tilde{w}h(\tilde{w})(1-\tilde{w}^2)^{3/2}}\dd\tilde{w}^2+\frac{4\mathcal{C}^2 h(\tilde{w})}{B}\dd \tilde{z}^2\bigg]\, ,
\ee
which is as given in \cite{Bah:2021hei}. Note that the gauge field and scalar are equivalent as well after the above redefinition. 
By using this dictionary one finds that the regularity analysis performed above agrees with the equivalent analysis performed in \cite{Bah:2021hei}.  

This concludes our review of the 7d solution and we turn our attention to the uplift of the solution to 11d supergravity on an $S^4$ .

%%%%%%%%%%%%%%%%%%%%%%%%%%%%%%%%%%%%%%%%%%%%
%%							11d solution
%%%%%%%%%%%%%%%%%%%%%%%%%%%%%%%%%%%%%%%%%%%%

\subsection{11d uplift and regularity}

In the previous section we have studied the regularity of the 7d solution. We have seen that the solution exhibits two distinct  singular behaviours; one at the centre of the disc and one along the boundary. In this section we will study the 11d uplift of the solution, focussing in particular on the singular regimes from the 7d solution. 

Using the uplifting formula in \cite{Cvetic:1999xp} the metric is
\begin{align}
\dd s^2_{11}
&=\Omega^{1/3}\big(w P(w)\big)^{1/5}\bigg[\dd s^2_7+\frac{1}{\Omega\big(w P(w)\big)^{1/5}}\Big( X_0^{-1}\dd \mu_0^2+\sum_{i=1}^{2} X_{i}^{-1}(\dd \mu_i^2+\mu_i^2(\dd \phi_i+A_{i})^2)\Big)\bigg]\, ,\label{eq:11dM5sol}
\end{align}
with 
\begin{align}
\Omega&=\sum_{I=0}^{2} X_I\mu_I^2\, ,\qquad X_0=X_1^{-2}X_2^{-2}\, , \qquad \sum_{I=0}^{2}\mu_I^{2}=1\, .
\end{align}
Note that $X_0=X_2$ for the disc. Given this symmetry it is useful to parametrise the $\mu_I$ as
\be
\mu_0=\sqrt{1-\mu^2}\cos\theta\, ,\quad \mu_2=\sqrt{1-\mu^2}\sin \theta\, ,\quad \mu_1=\mu\, ,
\ee
and to define
\be
\Omega=\frac{\tilde{\Omega}}{w^{4/5}h_1(w)^{3/5}}\, ,\qquad \text{with}\qquad \tilde{\Omega}=w^2 \mu^2+h_1(w)(1-\mu^2)\, .
\ee
Note that $\tilde{\Omega}$ vanishes at $(w=0, \mu=1)$ but is otherwise positive definite. Next define
\be
\hat{f}(w)=(w_--w)(w_+-w)\, ,
\ee
then the metric takes the form
\begin{align}
\dd s_{11}^2 =&~w^{1/3}\tilde{\Omega}^{1/3}\bigg[4 \dd s^2(\text{AdS}_5)+\frac{1}{w\hat{f}(w)}\dd w^2+\frac{\hat{f}(w)}{h_1(w)}\dd z^2 +\frac{w^2(1-\mu^2)}{w\tilde{\Omega}}\dd s^2(S^2)_{(\theta,\phi_2)}
\nonumber\\
&+\frac{h_1(w)\mu_1^2}{w \tilde{\Omega}}D\phi_1^2+\frac{1}{w(1-\mu_1^2)}\dd \mu_1^2\bigg]\label{eq:M511dmetricsimp}\, ,
\end{align}
with 
\be
h_1(w)=\hat{f}(w)+w\, ,\qquad D\phi_1=\dd\phi_1+A_1\, .
\ee
Note that the singularity at $w=0$ of the 7d metric persists in the uplifted solution. This is in contrast to the behaviour of the M2-brane and D3-brane disc solutions discussed in \cite{Couzens:2021rlk,Suh:2021hef} and \cite{Suh:2021ifj,Couzens:2021tnv} respectively, where the line $w=0$ is no longer singular only the point $(w=0, \mu=1)$.

\begin{figure}[h!]
\centering
  \includegraphics[width=0.6\linewidth]{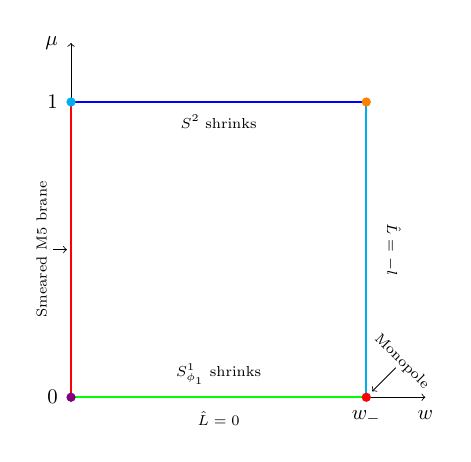}
  \captionsetup{width=.75\linewidth}
  \caption{A schematic plot of the rectangle over which the $S^2\times S^1_z\times S^1_{\phi_1}$ fibers are defined. In the interior of the rectangle all the fibers remain of finite size, whilst along the boundary various cycles shrink. The red dot on the bottom right hand corner is the location of a monopole of charge $l$. }
  \label{fig:rectangleM5}
\end{figure}

In order to interpret the solution it is useful to observe that it can be written in the form of an $S^2\times S^1_z\times S^1_{\phi_1}$ fibration over the rectangle $[0,w_-]\times [0,1]$. Away from the boundary of the rectangle the metric is smooth and the fibers are non-shrinking. Along the boundary various fibers shrink, see figure \ref{fig:rectangleM5}.\footnote{The rectangle we take looks somewhat different at first from the one in \cite{Bah:2021hei}. To compare the two one should take $\mu^2=1-\mu^2_{BBMN}$ }

Since the behaviour of the various edges of the rectangle will play a prominent role later we will study this in detail. The results we find are in agreement with the results in \cite{Bah:2021hei} and the reader familiar with the analysis there may skip to the next section safely. We will first study the degeneration along the sides away from the vertices. 
\\

\noindent {\bf $\mu=0$ degeneration}

\noindent 
Consider first the degeneration at $\mu=0$. We can see that the $\phi_1$ circle shrinks smoothly giving $\mathbb{R}^2$ if $\phi_1$ has period $2\pi$. This is of course the expected behaviour given the $S^4$ origin. \\

\noindent {\bf $\mu=1$ degeneration}

\noindent 
At $\mu=1$ we see that the $S^2$ shrinks smoothly giving $\mathbb{R}^3$. As before this is the expected result given the $S^4$ origin. 
\\

\vspace{3mm}
\noindent {\bf $w=w_-$ degeneration}

\noindent
To properly understand the degeneration at $w=w_-$ we should rewrite the metric in the form of the $S^1_z$ circle fibered over the $S^1_{\phi_1}$ circle. It is also convenient to perform a gauge transformation of the gauge field whilst performing the rewriting, $\delta A_1=n_z \dd z$. The choice of the constant $n_z$ we will make is such that the Killing spinor on the disc is independent of the angular coordinate. This is achieved by taking $n_z=-\tfrac{1}{2}$. The $S^1_z\times S^1_{\phi_1}$ part of the metric, ignoring the overall warp-factor $w^{1/3}\tilde{\Omega}^{1/3}$, after the rewriting takes the form
\be
\dd s^2(S^1_z\ltimes S^1_{\phi_1})=R_z^2 (\dd z+ L \dd \phi_1)^2+R_{\phi_1}^2 \dd \phi_1^2\, ,
\ee
with
\begin{align}
R_z^2&=\frac{ \mu^2(w^2-w_-w_+)^2 +4w \hat{f}(w)\tilde{\Omega}}{4w h_1(w) \tilde{\Omega}}\, ,\\
R_{\phi_1}^2&=\frac{4\mu^2 \hat{f}(w)h_1(w)}{\mu^2 (w^2-w_-w_+)^2 +4w \hat{f}(w)\tilde{\Omega}}\, ,\\
L&=\frac{2 \mu^2 (w^2-w_-w_+) h_1(w)}{ \mu^2(w^2-w_-w_+)^2+4w\hat{f}(w)\tilde{\Omega}}\, .
\end{align}
Note that $R_{\phi_1}$ vanishes at both $\mu=0$ and $w=w_-$ whilst $R_z$ only vanishes at the point $(w=w_-,\mu=0)$. Note also that the function $L$ is piecewise constant on the two-edges: $\mu=0$ and $w=w_-$. Along $\mu=0$ it vanishes, whilst along $w=w_-$ it is a non-zero constant. Since $L\dd \phi_1$ defines a connection for the fibration the physical parameter is
\be
\hat{L}(w,\mu)=\frac{2\pi L(w,\mu)}{\Delta z}\, ,
\ee
and we find $\hat{L}(w_-,\mu)=-l$. This signifies the presence of a monopole at $(w=w_-, \mu=0)$.

To see this more clearly let us take the simultaneous limit towards this point. We change coordinates to
\be
\mu=r \cos^2\Big(\frac{\zeta}{2}\Big)\, ,\qquad w=w_--\frac{w_+-w_-}{4}r^2 \sin^2\Big(\frac{\zeta}{2}\Big)\, .
\ee
and then take the $r\rightarrow 0$ limit. The $S^2$ metric remains of finite size and the remaining 4d part of the internal space becomes
\begin{align}
\dd s^2_4=\frac{1}{w_-}\bigg[\dd r^2+\frac{r^2}{4}\Big(\frac{4}{l^2}\big(   \dd \hat{z}- \tfrac{l}{2}(1+\cos\xi)\dd\phi_1\big)^2 +\dd \xi^2+\sin^2\xi \dd\phi_1^2\Big)\bigg]\, .
\end{align}
This is the metric on $\mathbb{R}^4/\mathbb{Z}_{l}$, and is due to the presence of a monopole.  
\\

\noindent {\bf $w=0$ degeneration}

\noindent
The final edge of the rectangle is the one along $w=0$. Series expanding along $w=0$ the metric reads
\begin{align}
\dd s^2\simeq\, &w^{1/3}\hat{f}(0)^{1/3}(1-\mu^2)^{1/3}\bigg[\Big(4\dd s^2(\text{AdS}_5)+\dd z^2\Big)+\frac{1}{w}\bigg\{\frac{1}{1-\mu^2}\dd \mu^2 +\frac{\mu^2}{1-\mu^2}\dd \phi_1^2 \nonumber\\
&+\frac{1}{\hat{f}(0)}\Big(\dd w^2 +w^2 \dd s^2(S^2)\Big)\bigg\}\bigg]\, .
\end{align}
Away from $\mu=1$ this is the metric on an M5-brane wrapping AdS$_5\times S^1_z$, located at the tip of $\mathbb{R}^3$ and smeared along the two directions spanned by $\mu$ and $\phi_1$.

%%%%%%%%%%%%%%%%%%%%%%%%%%%%%%%%%%%%%%%%%%%%
%%							Multi-Puncture
%%%%%%%%%%%%%%%%%%%%%%%%%%%%%%%%%%%%%%%%%%%%

\subsection{LLM reformulation}

In the previous section we have studied the uplifted solution. In this section we will extend this analysis by showing that the solution can be embedded into the classification of $\mathcal{N}=2$ preserving AdS$_5$ solutions of 11d supergravity of \cite{Gaiotto:2009gz,Lin:2004nb}.\footnote{In \cite{Bah:2021hei} they showed that the solution can indeed be embedded into the $\mathcal{N}=2$ classification. Since our conventions differ we will present the results from scratch as they are needed as an intermediate step.} Ultimately we want to consider the equivalent electrostatic description of the problem \cite{Gaiotto:2009gz}, but for ease of exposition we will present the intermediate steps. 
We will follow the conventions of \cite{Gaiotto:2009gz} in the following. Any $\mathcal{N}=2$ AdS$_5$ solution of 11d supergravity takes the following form:
\begin{align}
\dd s^2_{11}&=\,\me^{2\lambda}\bigg[4 \dd s^2(\text{AdS}_5)+ y^2 \me^{-6\lambda} \dd s^2(S^2)+\dd s^2_{4}\bigg]\, ,\label{eq:Toda1}\\
\dd s^2_4&= \frac{4}{1- y \partial_y D}D\chi^2-\frac{\partial_y D}{y}\Big(\dd y^2+\me^{D}(\dd x_1^2+\dd x_2^2)\Big)\, ,\label{eq:4dclass}\\
D\chi&=\dd \chi +v_i \dd x^i\, ,\qquad v_i=\frac{1}{2}\epsilon_{ij}\partial_{j} D\, ,\qquad v=v_i \dd x^i\\
\me^{-6 \lambda}&=-\frac{\partial_y D}{y(1-y \partial_y D)}\, ,\\
G_4 &= 2\dd \vol(S^2)\wedge \bigg[D\chi\wedge \dd (y^3 \me^{-6 \lambda})+y (1-y^2 \me^{-6 \lambda})\dd v -\frac{1}{2} \partial_y \me^{D} \dd x_1\wedge \dd x_2\bigg]\label{eq:Toda5}\, .
\end{align}
As before, the metric on AdS$_5$ is the unit radius one. The potential $D$, which determines the full solution is a solution of the (infinite) Toda equation
\be
\square_{x}D+\partial_y^2\me^{D}=0\, .\label{eq:Toda}
\ee

Comparing the general form of the metric with the solution 
\eqref{eq:M511dmetricsimp} we can immediately identify
\begin{align}
\me^{2 \lambda}=w^{1/3}\tilde{\Omega}^{1/3}\, .
\end{align}
After a little rewriting the metric takes the form
\begin{align}
\dd s^2_{11}&=\me^{2\lambda}\bigg[ 4 \dd s^{2}(\text{AdS}_5)+w^2(1-\mu^2)\me^{-6\lambda} \dd s^{2}(S^2)_{(\theta,\phi_2)}+\dd s^2_4\bigg]\, ,\\
\dd s^2_4&=\frac{\dd\mu ^2}{w(1-\mu^2)}+\frac{\mu_1^2 h_1(w)}{w\big((1-\mu^2)h_1(w)+\mu^2 w^2\big)}(\dd \phi_1+A_1)^2+ \frac{w}{f(w)}\dd w^2 +\frac{f(w)}{P(w)}\dz^2\dd \hat{z}^2\, .\nonumber
\end{align}
From the coefficient of the two-sphere we can then identify
\be
y=w \sqrt{1-\mu^2}\, .
\ee
Since the solution has an enhancement of symmetry compared to the general classification, there is an additional U$(1)$ symmetry, we define the polar coordinates
\be
x_{1}+\ii x_2=r \me^{\ii \beta}\, .
\ee
We now want to identify the radial coordinate $r$ and potential $D$ in terms of $w,\mu$. We find that they are given by
\begin{align}
r(w,\mu)&= H\bigg[\mu^2\big(w_+-w\big)^{\frac{2 w_+}{w_+-w_-}}\big(w_--w\big)^{-{\frac{2 w_-}{w_+-w_-}}}\bigg]\, ,\label{eq:rwithH}\\
\me^{D}&=\frac{w\mu}{\partial_w r(w,\mu)\partial_{\mu}r(w,\mu)}\, ,
\end{align}
with $H$ an arbitrary function of one variable with continuous first derivative.\footnote{The fact that the function $H$ is undetermined is due to the conformal symmetry of the solution in the Toda picture.} 
We can fix the function to be of the form $H(x)=x^{\alpha}$, with $\alpha$ a constant. There are different choices one could make for $\alpha$, for example if one takes $\alpha=\tfrac{1}{2}$ one finds that the potential $D$ is independent of the coordinate $\mu$. We will instead make the seemingly crazy choice $\alpha=-\dz^{-1}$, see equation \eqref{eq:deltaz} for the definition of $\dz$. This turns out to be useful because the metric takes the canonical LLM form upon making the change of coordinates
\be
\phi_1=-2\chi+\Big(2+\frac{\dz}{2}\Big)\beta\, ,\qquad \hat{z}=\frac{2}{\dz}\chi -\Big(1+\frac{2}{\dz}\Big)\beta\, ,
\ee
which importantly has Jacobian 1, leading to $2\pi$ periodic coordinates in the canonical $\mathcal{N}=2$ form. Other choices of $\alpha$ do not have this property.  
The potential and radial coordinate with this choice are\footnote{When $w_-=w_+$ the centre of the disc becomes $\mathbb{H}^2$ rather than $\mathbb{R}^2$. The radial coordinate is
\begin{equation*}
r=H[\mu^2\me^{2 w_-/(w_--w)}]\, .
\end{equation*} This behaviour is rather different to the more general case that we will study here. Note that at the $\mathbb{H}^2$ end-point the solution is of the form AdS$_5\times\mathbb{H}^2$. As we show in appendix \ref{app:MNlimit}, taking the limit carefully one obtains the Maldacena--Nunez solution. }
\be
r=\bigg[ \mu^2\big(w_+-w\big)^{\frac{2 w_+}{w_+-w_-}}\big(w_--w\big)^{-{\frac{2 w_-}{w_+-w_-}}}\bigg]^{-\frac{1}{\dz}}\, ,\qquad \me^{D}=\frac{\dz^2\mu^2 (w-w_-)(w-w_+)}{4  r(w,\mu)^2}\, .\label{eq:Dsol}
\ee
One can check that the potential satisfies the Toda equation \eqref{eq:Toda} as it should.

\subsection{Electrostatics reformulation}\label{sec:electro}

In the previous section we have rewritten the solution in terms of the classification of $\mathcal{N}=2$ AdS$_5$ solutions of 11d supergravity, determined by a potential satisfying the Toda equation. For solutions with two U$(1)$-isometries, like our solution, there is a formulation one can use by performing a B\"acklund transform \cite{Gaiotto:2009gz}. Rather than being determined by a potential satisfying the Toda equation, the solution is now determined by a potential satisfying the 3d cylindrical Laplace equation. In terms of this potential the problem can be interpreted as an electrostatic problem with a linear line-charge density $\lambda$ as we will now review.

To perform the B\"acklund transform we follow the conventions in \cite{Gaiotto:2009gz} and introduce the new coordinates $\rho,\eta$ defined via
\begin{align}
r^2 \me^{D}=\rho^2\, ,\quad y =\rho\partial_\rho V(\rho,\eta)\equiv \dot{V}\, ,\quad \log r=\partial_{\eta} V(\rho,\eta)\equiv V'\, .
\end{align}
The metric and flux after the coordinate transformation become
\begin{align}
\dd s^2=&\bigg(\frac{\dot{V}\tilde{\Delta}}{2 V''}\bigg)^{1/3}\bigg[ 4 \dd s^2 (\text{AdS}_5)+\frac{2 \dot{V}V''}{\tilde{\Delta}}\dd s^2(S^2)+\frac{2 V''}{\dot{V}}\Big(\dd \eta^2+\dd \rho^2+\frac{2 \dot{V}}{2\dot{V}-\ddot{V}}\rho^2 \dd \chi^2 \Big)\nonumber\\%
&+\frac{2(2 \dot{V}-\ddot{V})}{\dot{V}\tilde{\Delta}}\Big(\dd\beta +\frac{2 \dot{V}\dot{V}'}{2 \dot{V}-\ddot{V}}\dd \chi\Big)^2\bigg]\, ,\label{eq:electrometric}\\
C_3=&2\bigg[-\frac{2 \dot{V}^2 V''}{\tilde{\Delta}}\dd\chi+\Big(\frac{\dot{V}\dot{V}'}{\tilde{\Delta}}-\eta\Big)\dd\beta\bigg]\wedge \dd\vol(S^2)\, ,
\end{align}
where 
\begin{align}
\dot{\bullet}\equiv\rho\partial_{\rho}\, ,\qquad {\bullet}'\equiv \partial_{\eta}\, ,\qquad \tilde{\Delta}=(2 \dot{V}-\ddot{V})V''+(\dot{V}')^2\, .
\end{align}
The potential $V$ satisfies the 3d cylindrical Laplace equation
\be
\ddot{V}+\rho^2 V''=0\, .
\ee
To every potential giving rise to a sensible geometry and satisfying the Laplace equation one can define a line-charge density
\be
\lambda(\eta)=y(\rho=0,\eta)\, .
\ee
The benefit of the electrostatic description is that the cylindrical Laplace equation is linear and therefore we may construct more general solutions using superposition of known solutions.

Let us now turn our attention to obtaining the new coordinates $\rho, \eta$ and the potential $V$ for the disc solution. The coordinate $\rho$ is simple to extract in terms of $w,\mu$ and is given by
\be
\rho= \frac{\dz \mu\sqrt{(w_--w)(w_+-w)}}{2}\, ,\label{eq:rho}
\ee
where we have taken the positive root without loss of generality. To compute $\eta$ note that the integrability of the coordinate change implies the two constraints
\be
\partial_{\eta} y =\me^{D/2} \partial_{\rho}r\, ,\qquad \partial_\rho y =-\me^{D/2}\partial_{\eta} r\, ,
\ee
which are independent of the potential $V$. We may now solve for $\eta$ which gives
\be
\eta=\frac{\dz(1-2w)\sqrt{1-\mu^2}}{4}\, .
\ee
This is defined up to the addition of a constant, however since this may always be absorbed by a coordinate transformation later we set this constant to zero. With $\rho$ and $\eta$ in hand we can now determine the potential $V$,
\be
V=-\frac{1}{4}\bigg[\log \bigg(\frac{1+\sqrt{1-\mu^2}}{1-\sqrt{1-\mu^2}}\bigg)+\sqrt{1-\mu^2}\Big((1-2w)\log (r)-2\Big)\bigg]\, ,
\ee
which is also defined up to the addition of a constant, and again this constant is trivial. This indeed satisfies the cylindrically symmetric Laplace equation in 3d as it should.

We have now determined both the new coordinates $\rho,\,\eta$ and the potential $V$ for the electrostatic problem. However in our presentation above the potential is still written in terms of the original $w,\mu$ coordinates. To invert this we note that we may determine the $w,\mu$ coordinates in terms of $\rho,\eta$ as
 \begin{align}
w&=\frac{1}{4\dz}\bigg(2\dz- \nonumber\\
&\sqrt{2}\sqrt{\dz^2(1-4 w_- w_+)+16 (\eta^2+\rho^2)
+ \sqrt{\big(16(\eta^2+\rho^2)-\dz^2(w_+-w_-)^2\big)^2+64\dz^2 (w_+-w_-)^2 \rho^2}}\bigg)\, ,\nonumber\\ 
\mu^2&=\frac{4\rho^2}{\dz^2(w_--w)(w_+-w)}\, ,\label{eq:wmu}
\end{align}
where one should insert the expression for $w$ into $\mu^2$. The final potential is
\begin{align}
-4 V= \log \bigg( \frac{\dz(1-2 w)+4\eta}{\dz(1-2 w)-4\eta}\bigg) +\frac{4 \eta}{\dz} \log \Bigg[ \frac{4 \rho^2}{\dz^2} \bigg(\frac{w_{+}-w}{w_{-}-w}\bigg)^{\frac{1}{w_+-w_-}}\Bigg]-\frac{8 \eta}{\dz(1-2 w)}\, ,\label{eq:Vpot}
\end{align}
and $w$ should be understood to be the function of $\rho,\eta$, depending on the constants $p,l$ given in \eqref{eq:wmu}. It is interesting to note that the potential can be broken into three pieces each of which are solutions of the 3d cylindrical Laplace equation on their own: 
\begin{align}
-4 V_1&= \log\bigg( \frac{\dz(1-2 w)+4\eta}{\dz(1-2 w)-4\eta}\bigg)\, ,\nonumber\\
-4 V_2&=\frac{8 \eta}{\dz} \log\frac{2 \rho}{\dz}\, ,\\
-4 V_3&=-\frac{4\eta}{\dz} \bigg(\frac{2}{1-2w}-\frac{1}{w_+-w_-}\log\frac{w_+-w}{w_--w}\bigg)\, .\nonumber
\end{align}
The second term is the simplest, non-trivial solution to the cylindrical Laplace equation one can construct. Note that both pure AdS$_7$ and the Maldacena--Nunez solution have the same form of blocks, see appendix \ref{app:limits}. Of course this is expected given that both of these solutions can be obtained from the same local solution considered here as we show in appendix \ref{app:limits}. 
\\
%%%%%%%%%%%%%%%%%%%%%%%%%%%%%%%%%%%%%%%%%%%%
%%							Electrostatics
%%%%%%%%%%%%%%%%%%%%%%%%%%%%%%%%%%%%%%%%%%%%

\noindent {\bf Properties of the electrostatic setup}

\begin{figure}[h!]
\centering
  \includegraphics[width=0.5\linewidth]{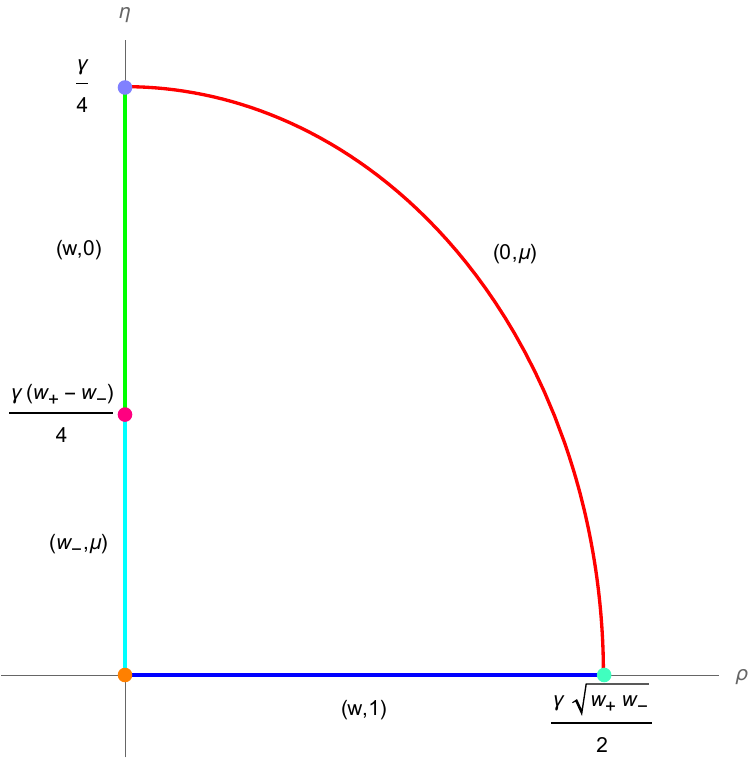}
  \captionsetup{width=.95\linewidth}
  \caption{{We plot the region in $(\rho,\eta)$ space where the electrostatic problem is defined. The coloured segments correspond to the different faces of the rectangle in $(w,\mu)$ space, see figure \ref{fig:rectangleM5}. The points are the vertices of this rectangle in $(w,\mu)$ space. Starting with the origin and going clockwise they are: \textcolor{orange}{orange=$(w_-,1)$}, \textcolor{magenta}{magenta=$(w_-,0)$ (monopole)}, \textcolor{violet}{violet=$(0,0)$}, \textcolor{cyan}{cyan=$(0,1)$}. The red line forming the ellipse is the location of the smeared brane, whilst the magenta dot (lying on the $\eta$-axis at the focus point of the ellipse) arises due to the presence of a monopole. }}
  \label{fig:rhoetadomain}
\end{figure}

\noindent Having reformulated the disc solution in terms of an electrostatic problem we will now study the solution from this vantage point. The first task is to identify the range of the $\rho$ and $\eta$ coordinates. By inserting the boundary values of $w,\mu$ we find that the boundary of the $(w,\mu)$-rectangle is identified with
\be 
\mu=0\Rightarrow \rho=0\, ,\quad \mu=1\Rightarrow \eta=0\, ,\quad w=w_-\Rightarrow \rho=0\, ,\quad w=0\Rightarrow \dz^2=16 \eta^2+\frac{4 \rho^2}{w_+w_-}\, .\label{eq:ranges}
\ee
We find that the ranges of $\rho$ and $\eta$ are
\be
\rho\in \Big[0,\frac{\dz\sqrt{w_+ w_-}}{2}\Big]\, ,\qquad \eta \in \Big[0,\frac{\dz}{4}\Big]\, .
\ee
Note that the location of the smeared branes (irregular puncture) defines an ellipse in the $(\rho,\eta)$ coordinates, see the last condition in \eqref{eq:ranges}.
The focal point of the ellipse is at 
\be
(\rho,\eta)_{\text{Focus}}=\Big(0,\pm \dz\frac{w_+-w_-}{4}\Big)\, ,
\ee
which, for the sign within the domain ($+$), is the location of the monopole! In figure \ref{fig:rhoetadomain} we have plotted the resultant domain in the $(\rho,\eta)$ coordinates, colour coded to match the regions in figure \ref{fig:rectangleM5} for the $(w,\mu)$ coordinates.

The prototypical example of solutions in the literature of this electrostatic problem includes a non-compact domain, with both $\rho$ and $\eta$ non-compact, see for example \cite{Gaiotto:2009gz,Lin:2004nb}. The typical boundary condition imposed is that $\dot{V}$ vanishes along $\eta=0$, and we find that this is also true of the solutions discussed here. Some solutions with $\eta$ compact have been found in the literature, see for example \cite{Lozano:2016kum,Aharony:2012tz,Reid-Edwards:2010vpm}, where the domain is a rectangle. The additional boundary condition imposed in these works is that $\dot{V}$ vanishes along a line $\eta=\eta_*$ for some $\eta_*>0$.
Our compact domain is quite novel being a quarter of an ellipse. The inclusion of the smeared brane (irregular puncture) leads to a non-trivial boundary condition for the electrostatic problem, that is the boundary condition that $\dot{V}$ vanishes on an ellipse. This sets the disc solution apart from previous examples in the literature.\\

\noindent {\bf Line Charge}

\noindent Having determined the potential $V$ it is a simple matter to obtain the line charge $\lambda(\eta)$. We find 
\be
\lambda(\eta)=y(\rho=0,\eta)=\begin{cases}
\frac{l p}{p+1}\eta & \eta\in [0, \tfrac{1}{2 l}]\, ,\\
\frac{1}{2}-\frac{l}{p+1}\eta & \eta \in [\tfrac{1}{2 l},\tfrac{p+1}{2 l}]\, ,
\end{cases}\label{eq:linechargev1}
\ee
where we have used \eqref{eq:rootsp} and \eqref{eq:deltaz} to express the result in terms of the magnetic charge defined in \eqref{eq:defp} and the orbifold weight. Note that the change in slope at the monopole point is $l$, the orbifold weight. 
The reader familiar with the conditions in \cite{Gaiotto:2009gz} may be uneasy that $\eta$ does not take integer values at the monopole point and that the slope is not integer. As we will explain later this still gives rise to a well-defined solution, in fact the constraints in \cite{Gaiotto:2009gz} are too strong and not all constraints imposed there are needed for a well-defined solution of this type. As we will see the non-integer slope leads to operators in the dual field theory having non-integer scaling dimensions. 

As an aside, note that there is a scaling symmetry of the solution. One may perform the transformation
\be
V(\rho,\eta,p)\rightarrow M ~V(\lsc \rho,\lsc \eta+c,p)\, ,\label{eq:Vscaling}
\ee
for $M$ and $\lsc$ some constants, and retain a solution to the cylindrical Laplace equation. This transformation leads to a transformation of the line charge as
\be
\lambda(\eta)\rightarrow M \lambda(\lsc \eta +c)\,,
\ee
and therefore one could use this freedom to make the line charge satisfy the conditions in \cite{Gaiotto:2009gz}. Note that the parameter $c$ is precisely the linear shift we could have performed earlier when obtaining the coordinate $\eta$. 
We will refrain from performing these rescalings for the time being. \\

\noindent {\bf Line charge kinks analysis}

\noindent  Let us now substantiate our claim that the following line charge and potential do give rise to a well-defined geometry. We will show that the monopole number is indeed $l$ and that the flux is properly quantised. Since the analysis goes through allowing for an arbitrary number of kinks in the line charge and we will need this later, we perform the analysis allowing for $f$ kinks in the line charge. We perform the analysis by assuming that the line-charge and potential $V$ have the properties of the disc solution, in particular that $\dot{V}$ vanishes along $\eta=0$ and along some curve generalising the ellipse above. The final results will depend only on the implicit line-charge and not the details of this curve. As such let us denote the locations of the kinks of the line charge to be $n_a$ with $a=\{1,..,f\}$ and let the changes in slope be $l_{a}$. Finally the end-point of the line charge, where $\lambda(\eta)$ vanishes will be taken to be $n_{f+1}$. 

First let us consider the expansion of the metric around a kink of the line charge, taking this point to be $n$. 
Let us perform the change of coordinates
\be
\rho=r^2 \sin\zeta\, ,\qquad \eta=n+r^2\cos\zeta\, , \qquad \zeta\in [0,\pi]\, ,
\ee
which form semi-circles around the monopole in this 2d plane. 
Expanding the terms appearing in the metric around $r=0$ we have
\begin{align}
\dot{V}&=\lambda(n)\,, \quad \ddot{V}=\frac{r^2\sin\zeta}{2}\delta\lambda(n)\, ,\quad V''=-\frac{1}{2 r^2}\delta\lambda(n)\, ,\quad \nonumber\\
\dot{V}'&=\frac{1}{2}\delta\lambda(n) (\cos\zeta-1)-\sum_{m>n}\delta\lambda(m)\, ,\quad \Delta=-\frac{1}{r^2}\lambda(n)\delta\lambda(n)\, ,\label{eq:monopoleexpansion}
\end{align}
where
in taking the derivative of $\lambda$ around the monopole point we have
\be
\delta\lambda(n)=\lim_{\epsilon\rightarrow 0^{+}}\big(\lambda'(n+\epsilon)-\lambda'(n-\epsilon)\big)\, ,
\ee
and the sum is over monopoles higher along the $\eta$ axis.  
In this limit the metric becomes
\begin{align}
\dd s^2&=\lambda(n)^{2/3}\bigg[4 \dd s^2(\text{AdS}_5)+\dd s^2(S^2)\\
&-\frac{4\delta \lambda(n)}{\lambda(n)}\bigg\{ \dd r^2+\frac{r^2}{4}\Big(\dd\zeta^2+\sin^2\zeta\dd\chi^2
+4\Big(|\delta\lambda(n)|^{-1}\dd\beta+\big(\tfrac{1}{2}(\cos\zeta-1)+\alpha\big)\dd\phi\Big)^2\Big)\bigg\}\bigg]\, ,\nonumber
\end{align}
where $\alpha$
is an additive constant which can be removed by a gauge transformation. This is then the metric on AdS$_5\times S^2\times \mathbb{R}^4/|\delta\lambda(n)|$ and implies that we should take $\delta\lambda(n)$ to be a negative integer. This imposes that the line charge density is convex and has integer changes in the slope at a location of the monopole. \\

\noindent {\bf Flux quantisation}

\noindent  Next consider flux quantisation. We may rewrite the three-form potential in the form
\be
C_3=\bigg[f_\chi(\rho,\eta)\dd\chi+f_{\beta}(\rho,\eta)\dd\beta\bigg]\wedge \dd\vol(S^2)\, ,
\ee
with
\be
f_{\chi}(\rho,\eta)=-\frac{4 \dot{V}^2V''}{\tilde{\Delta}}\, ,\qquad f_{\beta}(\rho,\eta)=2\bigg(\frac{\dot{V}\dot{V}'}{\tilde{\Delta}}-\eta\bigg)\, .
\ee
Flux quantisation imposes that
\be
\frac{1}{(2\pi\lp)^3}\int_{\Sigma}G_4\in \mathbb{Z}\, ,
\ee
for all integral four-cycles $\Sigma$. 

We must first identify all integral four-cycles in the geometry. There are two types of four-cycle to consider depending on shrinking cycles in the geometry. The first type of cycle, which we denote by $\mathcal{C}_a$, are constructed by taking the cycle which shrinks along $\rho=0$ and the two-sphere which shrinks along $\eta=0$ which is topologically a four-sphere. Pictorially they are given by a line stretching from a point on the $\eta$ axis to the $\rho$ axis. For $f$ kinks there are $f+1$ such cycles that can be constructed by taking the shrinking cycle to be located at some point between two kinks $n_{a}$ (the end-points included). 

The second class of four-cycle is topologically either $S^2\times S^2$ or $S^4$. and is constructed by taking the shrinking $\beta$ cycle along $\rho=0$ between two kinks (or the end-points) and the round two-sphere, and will be denoted by $\mathcal{D}_a$. For the cycles involving the two end-points the four-cycle is topologically an $S^4$ and otherwise is topologically $S^2\times S^2$.

Let us first consider the four-cycles $\mathcal{C}_a$, and let $a=\{0,...,f\}$. We must identify the exact $S^1$ that is shrinking along $\rho=0$, this depends where it is located along the $\eta$-axis. We may identify the $2\pi$ periodic coordinate which parametrises this shrinking $S^1$ by finding a Killing vector with vanishing norm as $\rho\rightarrow 0$ which is normalised such that the surface gravity is $1$. The Killing vector satisfying these properties is\footnote{When $\lambda(\eta)'$ is integer in all intervals along $\rho=0$ the two U$(1)$'s give rise to a regular fibration of a $\mathbb{T}^2$, however if $\lambda(\eta)'$ is instead rational but not integer, then this gives rise to a quasi-regular fibration. As we will see, there is no requirement from imposing regularity conditions on the metric nor flux quantisation that constrains $\lambda(\eta)'$ to be integer generically, this is an assumption made in \cite{Gaiotto:2009gz}. What is constrained is the difference of $\lambda(\eta)'$ between segments which must be integer. As we will see the disc solution gives rise to a quasi-regular fibration generically and $\lambda(\eta)'$ is not integer, though the difference between segments is. The assumption that $\lambda(\eta)'$ is integer is useful for writing the explicit quiver but is not essential as we will explain at the end of this section. When considering a line-charge which plateaus rather than vanishes one must take the slopes to be integer however.  }
\be
\partial_{\varphi}=\partial_{\chi}-\lambda'(\eta) \partial_{\beta}\, .
\ee
Note that since $\lambda(\eta)$ is linear that this is independent of the coordinate $\eta$, but depends on the location along the $\rho=0$ line due to the jumps in $\lambda'(\eta)$.
We find
\begin{align}
\frac{1}{(2\pi\lp)^3}\int_{\mathcal{C}_a}G_4 &=\frac{8\pi}{(2\pi \lp)^3}\Big[f_{\chi}(\rho,\eta)-\lambda'(\eta)f_{\beta}(\rho,\eta)\Big]_{\eta=0}^{\rho=0}\nonumber\\
&=-\frac{16 \pi^2}{(2\pi\lp)^3}~\lambda(\eta)\Big|_{\text{constant piece}}\, ,
\end{align}
where we take $\lambda$ in the interval given by the index `$a$'. This should be integer for all four-cycles. Note that since $\lambda(0)=0$ the flux through the cycle $\mathcal{C}_0$ is 0 always, whilst the cycle $\mathcal{C}_f$ gives the total number of M5-branes wrapping the punctured sphere. 

Consider now the second type of cycle $\mathcal{D}_a$. Let the locations of the monopole along the $\eta$ axis be denoted by $n_a$, with $n_0$=0, and $n_{f+1}$ the end-point of the $\eta$-axis. We have
\begin{align}
\frac{1}{(2\pi \lp)^3}\int_{\mathcal{D}_a}G_4&=-\frac{16\pi^2}{(2\pi \lp)^3}(n_{a}-n_{a-1})\, .
\end{align}

To simplify the results let us take the line charge to be the union of lines of the schematic form
\be
\lambda= r_a \eta +m_a\, .
\ee
Then the quantisation conditions in general are equivalent to
\begin{align}
r_{a-1}-r_{a}\equiv l_a\in\mathbb{Z}\, ,\qquad \frac{16 \pi^2}{(2\pi\lp)^3} m_a\equiv M_a\in \mathbb{Z}\, ,\qquad \frac{16\pi^2}{(2\pi \lp)^3} n_{a}\equiv N_a\in\mathbb{Z}
\end{align}
We should understand $M_a$ as the number of M5-branes making the puncture with $M_{f}$ the total number of M5-branes. 

For the line charge in \eqref{eq:linechargev1} we have
\be
n_0=0\, ,\quad n_1=\frac{1}{2l}\, ,\quad n_2=\frac{p+1}{2l}\, ,\quad r_0=\frac{l p}{p+1}\, ,\quad r_1=-\frac{l}{p+1}\, ,\quad m_0=0\, ,\quad m_1=\frac{1}{2}\, .
\ee
Note that $r_0-r_1=l$ and is integer. The quantisation conditions are satisfied if
\be
\frac{1}{\pi l \lp^3}=N\in \mathbb{Z}\, ,
\ee
and $p\in \mathbb{Z}$. Note that this is different from the quantisation condition considered in \cite{Gaiotto:2009gz}, where the $l$ is not present. We emphasise that despite the line-charge not satisfying all the properties required in \cite{Gaiotto:2009gz}, in particular the integer slopes the solution is globally well-defined. In \cite{Gaiotto:2009gz} for certain choices of line charge where it plateaus one is still forced into having integer slopes, however when the line charge does not plateau and either increases to infinity or has a zero away from $\eta=0$ then this assumption is too restrictive. As we will see when considering the field theory, the non-integer slope leads to operators in the dual SCFT with fractional scaling dimensions. 

Having understood the constraints for a well-defined solution let us consider the scaling symmetry \eqref{eq:Vscaling} which we may use to redefine the parameters $n_a$. 
Under this scaling symmetry the parameters labelling the solution transform as 
\be
n_a\rightarrow \frac{n_a}{\lsc}\, ,\quad r_a\rightarrow M \lsc r_a\, ,\quad m_a\rightarrow M m_a\, .
\ee
We want to keep the change in the slope at the kinks fixed under the symmetry since this gives rise to the physical parameter labelling the orbifold weight. This forces us to take $M \lsc=1$ in the following and it follows that the scaling symmetry acts the same on $m_a$ and $n_a$. Consequently, we may absorb this scaling by redefining the flux quanta $N$: this is to be expected since this scaling symmetry, whilst holding the change in the slope at the kink fixed, does not give rise to a physical parameter. We can then use the symmetry to fix $M=\lsc^{-1}= 2 l$ which makes the new distinguished locations, $n_a$, integer. We find the rescaled variables
\be
n_0=0\, ,\quad n_1=1\, ,\quad n_2=p+1\, ,\quad m_0=0\, ,\quad m_1=l\, , \quad N=\frac{2}{\pi \lp^3}\, .\label{eq:rescaledN}
\ee
We should emphasise that this rescaling has not changed the physics in any way since it can be absorbed in the relation between $N$ and the Planck length $\lp$. The parameter $N m_2$ gives the total number of M5-branes wrapped on the two-sphere whilst the flux $N m_1$ measures the number of M5-branes giving rise to the regular puncture at $N n_1$. 
Note the identity
\be
(r_0-r_1) n_1=-r_1 n_2\, ,
\ee
which follows since line-charge vanishes at two points. More generally if there are a total of $f$ kinks we have
\be
\sum_{a=1}^{f} N_a(r_{a-1}-r_{a})= - r_{f} N_{f+1}\, .
\ee 
Observe that the convex condition implies that each term in the sum on the left-hand-side is positive and consistency requires each term to be integer, thus the right-hand-side is also positive and integer. We can interpret this constraint as a partitioning of the positive integer $-  r_f N_{f+1}$ and will be related to the construction of a Young diagram for the regular puncture in section \ref{sec:fieldtheory}.
In \cite{Bah:2021hei} it was shown that this particular gravity solution is dual to a certain Argyres--Douglas theory. The goal of this paper is to generalise this construction. 

%%%%%%%%%~~~~~~~~~~~~~~~~~~~~~~~~~~~~~~~~~~~~~~~~~~~~~~~~~~~~~~~~~~
%%						Central Charge
%%%%%%%%%~~~~~~~~~~~~~~~~~~~~~~~~~~~~~~~~~~~~~~~~~~~~~~~~~~~~~~~~~~

\subsection{Observables}

Having quantised the fluxes and checked regularity let us turn our attention to the observables of the theory which we can compare to the field theory. One such observable is the `$a$' central charge, which we study in the following section. A second is the scaling dimension of probe M2-branes, and the third observable that we will consider is the flavour central charges of the solution.\\

\noindent {\bf Central Charge }

\noindent 
The leading order contribution to the $a$ central charge for an AdS$_5$ solution of the form \cite{Gauntlett:2006ai}
\be
\dd s^2_{11}= \me^{2 A}\Big[4\dd s^2(\text{AdS}_5)+\dd s^2(M_6)\Big]\, ,\label{eq:genmet}
\ee
is
\be
a=\frac{2^5\pi^3}{(2\pi \lp)^9} \int_{M_6}\me^{9 A} \dd \vol(M_6)\, .
\ee
From the form of the metric in electrostatic coordinates we identify
\begin{align}
\me^{9A}&=\Big(\frac{\dot{V}\tilde{\Delta}}{2V''}\Big)^{3/2}\, ,\\
\dd\vol(M_6)&= \frac{8 \sqrt{2}\rho (V'')^{5/2}}{\dot{V}^{1/2}\tilde{\Delta}^{3/2}}\dd \vol(S^2)\wedge \dd \eta\wedge\dd\rho\wedge \dd \chi\wedge \dd \beta\, ,
\end{align}
and therefore we have
\be
a= \frac{2^7\pi^3}{(2\pi \lp)^9} \int_{M_6}\rho\dot{V}V'' \dd \vol(S^2)\wedge \dd \eta\wedge\dd\rho\wedge \dd \chi\wedge \dd \beta\, .
\ee
Using the cylindrical Laplace equation we may rewrite this as 
\be
a=\frac{2^{10} \pi^6}{(2\pi \lp)^9}\int \partial_{\rho}(\dot{V}^2)\dd\eta\wedge \dd\rho\, .
\ee
It follows that we can integrate over $\rho$ by defining the integration domain to go from $\rho=0$ to the ellipse 
\be
n^2(p+1)^2 =\eta^2+\frac{(p+1)^2}{p(p+2)}\rho^2\, ,
\ee
and gives
\be
a=\frac{2^{10} \pi^6}{(2\pi\lp)^9}\int\bigg[\lambda(\eta)^2- \dot{V}(\rho,\eta)^2\Big|_{\text{ellipse}}\bigg]\dd \eta\, ,
\ee
where it is understood that the second term is evaluated on the ellipse by eliminating the $\rho$ coordinate. 
However, the boundary condition along the ellipse requires $\dot{V}$ to vanish and therefore the only contribution comes from the line charge term and we have\footnote{A similar expression for the central charge appears in \cite{Nunez:2019gbg}.}
\be
a=\frac{2^{10} \pi^6}{(2\pi\lp)^9}\int\lambda(\eta)^2\dd\eta\, .\label{eq:agen}
\ee
Note that this holds generally for a solution whose domain is fixed like the disc solution, i.e. has a similar boundary structure to figure \ref{fig:rhoetadomain} where the ellipse may be replaced by a more complicated curve along which $\dot{V}=0$. 
Carefully performing the integral we find
\be
a=\Big(\frac{2}{\pi \lp^3}\Big)^3\frac{l^2 p^2}{12(p+1)}\, ,
\ee
which upon using the quantisation condition \eqref{eq:rescaledN} gives
\be
a=N^3 \frac{ l^2 p^2}{12(p+1)}\, .
\ee
This agrees with the result in \cite{Bah:2021hei} upon using the following dictionary between our variables
\be
N_{BBMN}=lN\, ,\quad K_{BBMN}=p N\, .\label{eq:BBMNparadictionary}
\ee

\vspace{4mm}

%%%%%%%%%~~~~~~~~~~~~~~~~~~~~~~~~~~~~~~~~~~~~~~~~~~~~~~~~~~~~~~~~~~
%%						Scaling Dimensions
%%%%%%%%%~~~~~~~~~~~~~~~~~~~~~~~~~~~~~~~~~~~~~~~~~~~~~~~~~~~~~~~~~~

\noindent {\bf Scaling dimensions of probe M2-branes}

\noindent  One may wrap probe M2-branes around calibrated two-cycles in the geometry giving a BPS particle. With the metric in \eqref{eq:genmet} the calibration condition on a 2d submanifold $\Sigma_2$ reads \cite{Gauntlett:2006ai} 
\be
X\Big|_{\Sigma_2}=\dd \vol_{M_6}(\Sigma_2)\, ,
\ee
where $X$ is the calibrated two-form which can be constructed as a spinor bilinear and will be given momentarily. The right-hand-side denotes the restriction of the volume form on $M_6$ to the 2d submanifold. For such a calibrated two-cycle the conformal dimension of the BPS particle is given by
\be
\Delta=\frac{4\pi}{(2\pi\lp)^3}\int_{\Sigma_2}\me^{3 A} X\, .
\ee
The calibration two-form $X$ was given in \cite{Bah:2021hei,Lin:2004nb} in terms of the Toda frame and we refer the reader to \cite{Bah:2021hei} in particular for further details on its construction. Following \cite{Bah:2021hei} it is convenient to write the metric on the round two-sphere as
\begin{align}
\dd s^2(S^2)=\frac{\dd\tau^2}{1-\tau^2}+(1-\tau^2)\dd\varphi^2\, ,
\end{align}
with $\tau\in [-1,1]$, $1$ giving the north pole and $-1$ the south pole,  and $\varphi$ is $2\pi$-periodic. Then, the calibration two-form $X$ in the Toda frame (see \eqref{eq:Toda1}-\eqref{eq:Toda5}) is\footnote{There are some factors of 2 and signs different between the expression here and the one presented in \cite{Bah:2021hei} which are related to the different normalisation we employ and the different choices of volume form.}
\begin{align}
X&=y^3 \me^{-9 \lambda}\dd\vol(S^2)+y \me^{-3 \lambda}(1-y^2 \me^{-6 \lambda}) \dd\tau \wedge D\chi -\tau \me^{-3 \lambda}D\chi\wedge \dd y+\frac{\tau y \me^{-9 \lambda} \me^{D}}{1-y^2 \me^{-6 \lambda}}\dd x_1\wedge \dd x_2\, .
\end{align}
We may transform this to the electrostatic description by the change of coordinates
\begin{align}
y=\dot{V}\, ,\quad x_1+\ii x_2= r\me^{\ii \beta}\, ,\quad \log r=V'\, ,\quad r^2 \me^{D}=\rho^2\, ,\quad \beta\rightarrow \beta\, ,\quad \chi\rightarrow \chi+\beta\, ,
\end{align}
which gives the calibration two-form 
\begin{align}
X&=2\Big(\frac{2 V''}{\dot{V}\tilde{\Delta}}\Big)^{1/2}\bigg[ \frac{\dot{V}^2 V''}{\tilde{\Delta}}\dd\vol (S^2) +\tau\Big(\dot{V}'\dd\eta+\rho^{-1}\ddot{V}\dd\rho\Big)\wedge \Big(\dd\chi+\frac{\dot{V}'}{(\dot{V}')^2-\ddot{V}V''}\dd\beta\Big)\\
&
+\frac{\dot{V}}{\tilde{\Delta}}\dd\tau\wedge \Big(\big((\dot{V}')^2-V''\ddot{V}\big)\dd\chi+\dot{V}' \dd\beta\Big)
+\frac{\tau\rho^2 V''}{(\dot{V}')^2-\ddot{V}V''}\big(V'' \dd\eta+\rho^{-1}\dot{V}' \dd\rho\big)\wedge \dd \beta\bigg]\, .\nonumber
\end{align}
Note that this holds for any solution written in electrostatic coordinates following the conventions used in this paper. 
The calibrated two-cycle that we will consider in the following is the round two-sphere at fixed positions in $M_6$. The calibration condition forces the cycle to be located at the positions of the kink of the line charge. 

Let us check the calibration condition for the two-cycles. The calibration condition is
\be
2\dot{V} V^{'' }= \tilde{\Delta}\, .
\ee
This condition holds at the monopole point as one can verify from the expansion in \eqref{eq:monopoleexpansion}, in fact this is the only point that it holds true for the potential we study. 
The conformal dimension of the BPS particle is then
\begin{align}
\Delta(\mathcal{O}_1)&=\frac{4\pi}{(2\pi\lp)^3}\int_{S^2}\lambda(n)\dd \vol(S^2)\nonumber\\
&=\Big(\frac{2}{\pi \lp^3}\Big) \lambda(n)\, .\label{eq:dim1}
\end{align}
For the solution in BBMN we find
\be
\Delta(\mathcal{O}_1)=N \frac{l p}{p+1}\, ,
\ee
which agrees with the result in \cite{Bah:2021hei} upon using the dictionary provided in equation \eqref{eq:BBMNparadictionary}.\\

\noindent {\bf Flavour symmetries}

\noindent To each flavour symmetry we can associate a flavour central charge. One can use anomaly inflow methods to compute a mixed U$(1)_R$-flavour Chern--Simons term in AdS$_5$. From \cite{Gaiotto:2009gz} we have that the contribution to the flavour central charge due to the flavour group at the kink is
\be
k(\text{SU}(l))=\Big(\frac{4}{\pi\lp^3}\Big)\lambda(n)\, ,\label{eq:flavourgrav}
\ee
which gives
\be
k(\text{SU}(l))=2N \frac{ l p}{p+1}\, ,
\ee
in agreement with \cite{Bah:2021hei} after using the dictionary in equation \eqref{eq:BBMNparadictionary}.
As a consistency check we have obtained this result from an anomaly inflow computation in appendix \ref{app:anom}. 

%%%%%%%%%%%%%%%%%%%%%%%%%%%%%%%%%%%%%%%%%%%%
%%							Multi-Puncture
%%%%%%%%%%%%%%%%%%%%%%%%%%%%%%%%%%%%%%%%%%%%

\section{Generalised regular puncture }\label{sec:multi}

In the previous section we have discussed how to rewrite the disc solution of \cite{Bah:2021hei} in terms of an electrostatic problem. We have obtained a potential $V$ which depends on two integers, $p$ and $l$. Since the cylindrical Laplace equation is linear we may sum different potentials and obtain a new solution. In particular we can sum up the potential of the previous section with the different potentials depending on different parameters $p$ and $l$. We will interpret this as giving rise to a general Young diagram for the regular puncture. As we will show there are some constraints that the potential blocks must satisfy in order to give rise to a consistent solution. In the following sections we will analyse this in detail and compute the observables of the theory considered in the previous section.

\subsection{Building block of the generalised regular puncture}

In this section we will construct the general building block form of the potential which we will use in the remainder of the section. 
We saw earlier that the potential directly following from the disc solution can be more conveniently written by using the constant scaling symmetry of the electrostatic description. In order to make the discussion of the superposition of potentials simpler we will use this symmetry to construct the building block potential. Taking the potential in \eqref{eq:Vpot} and transforming it via
\be
V(\rho,\eta;\, p,l)\rightarrow  2 l n V\Big(\frac{\rho}{2 l n},\frac{\eta}{2ln};p,l\Big)=\mathcal{V}(\rho,\eta;p,l,n)
\ee
we end up with the building block 
\begin{align}
\mathcal{V}(\rho,\eta;\, p,l,n)&=\frac{l}{2}\bigg\{\frac{2 \eta}{(p+1)(1-2 w)}-n \log\Big[\frac{n(p+1)(1-2w)+\eta }{n(p+1)(1-2w)-\eta}\Big]-\frac{2\eta}{p+1} \log\frac{\rho}{2 n(p+1)}\nonumber\\
& -\eta \log\Big[\frac{2+p-2(p+1) w}{p-2(p+1) w}\Big] \bigg\}\label{eq:potential1}
\end{align}
where
\be
w(\rho,\eta;\, p,n)=\frac{1}{2}-\frac{1}{2\sqrt{2} n(p+1)}\sqrt{n^2+\eta^2+\rho^2+\sqrt{(\rho^2+\eta^2-n^2)^2+4 n^2 \rho^2}}\, .\label{eq:wdef}
\ee
Note that we have included an additional integer parameter $n$ which places the kink at $n$ rather than at 1. This parameter is trivial in the single kink case since it could be absorbed by redefining the flux quanta $N$, however, with multiple kinks this is no longer true and it becomes a bonafide parameter. The domain for the building block potential is a quarter of an ellipse satisfying
\be
\rho\in [0,n \sqrt{p(p+2)}]\, ,\quad \eta \in [0,n(p+1)]\, ,\quad 1\geq \frac{\eta^2}{n^2(p+1)^2}+\frac{\rho^2}{n^2p (p+2)}\, ,\label{eq:domainVblock}
\ee
with focus at 
\be
\rho=0, \,\quad \eta=n \, .
\ee
It is interesting to note that the level-sets of $w$ are ellipses defined by
\be
w(\rho,\eta;\, p,n)=\frac{\alpha}{1+2\alpha}\, ,\quad \Leftrightarrow \quad \frac{n^2}{(1+2\alpha)^2}=\frac{\eta^2}{(p+1)^2}+\frac{\rho^2}{(p-2\alpha)(p+2(\alpha+1))}\, ,\label{eq:levelsets}
\ee
with focus at $n$. The maximal value of the level set is 
\be
w_{\text{max}}=\frac{p}{2(p+1)}\, ,
\ee
which is of course the value of the root of the disc, and occurs at $\rho=\eta=0$. In terms of $\alpha$ this corresponds to $2\alpha=p$ and therefore we should take $-1\leq 2\alpha\leq p$.

We may rewrite the potential block $\mathcal{V}$ by first defining 
\be
\hat{w}(\rho,\eta;\,n)=\frac{1}{\sqrt{2}n}\sqrt{n^2+\eta^2+\rho^2+\sqrt{(\rho^2+\eta^2-n^2)^2+4 n^2 \rho^2}}\, ,
\ee
then
\begin{align}
\mathcal{V}(\rho,\eta;\,p,l,n)&=\frac{l}{2}\bigg\{\frac{2 \eta}{\hat{w}(\rho,\eta;n)}-\eta \log\Big[\frac{\hat{w}(\rho,\eta;n)+1}{\hat{w}(\rho,\eta,n)-1}\Big]-n 
\log\Big[\frac{n \hat{w}(\rho,\eta;n)+\eta}{n\hat{w}(\rho,\eta;n)-\eta}\Big]-\frac{2\eta}{p+1}\log\rho\bigg\}\, ,\label{eq:Vpotfinal}
\end{align}
and we have removed the following term 
\be
\delta \mathcal{V}=\frac{2\eta}{p+1}\log [2 n (p+1)]\, ,
\ee
from the original potential in \eqref{eq:potential1} since it is a trivial term and does not appear in any of the functions describing the solution as it is linear in $\eta$.
Note that the potential satisfies
\be
\rho\partial_{\rho}\mathcal{V}=\frac{2 l\eta }{p+1} \frac{w(\rho,\eta;\, p,n)}{1-2w(\rho,\eta;\, p,n)}=\frac{l \eta}{p+1}\frac{p+1-\hat{w}(\rho,\eta;n)}{\hat{w}(\rho,\eta;n)}\, ,
\ee
which vanishes when either $\eta=0$ or $w=0$ ($\hat{w}=p+1$). The latter has a line of zeroes on the ellipse defined in \eqref{eq:domainVblock}.  
The line-charge for the potential is
\be
\lambda(\eta)=\begin{cases}
\frac{p l}{p+1}\eta\, ,& 0\leq \eta\leq n\, ,\\
n l-\frac{l}{p+1}\eta\, ,& n\leq \eta \leq n(p+1)\, .
\end{cases}
\ee

%%%%%%%%%%%%%%%%%%%%%%%%%%%%%%%%%%%%%%%%%%%%
%%							Summing potentials
%%%%%%%%%%%%%%%%%%%%%%%%%%%%%%%%%%%%%%%%%%%%

\subsection{Summing up building blocks}

We now want to sum an arbitrary number of the building block potentials $\mathcal{V}(\rho,\eta;\,p,l,n)$ with different integers $p,l,n$. We must take the $l$'s integer and we choose to take the $n$'s integer too for later simplicity, this can always be arranged by the quantisation condition. The $p$'s are no longer constrained to be integer though. 
 Let us index the different potentials by a subscript $a$ with $a=\{1,...,f+1\}$, and order the $n_a$ so that $0<n_1< n_2<...< n_f<n_{f+1}$. The largest of the $n$'s, $n_{f+1}$ is the end-point and we understand $p_{f+1}=l_{f+1}=0$ and $n_0=0$. Then, the general potential is
\be
V=\sum_{a=1}^{f}\mathcal{V}(\rho,\eta;\, p_a,l_a,n_a)\, ,
\ee
and depends (naively) on $3f$ parameters in total. However, as can be seen from the form of the potential in \eqref{eq:Vpotfinal} the final potential actually depends only on $2f+1$ parameters: the $f$ $l_a$'s, the $f$ $n_a$'s and the end-point $n_{f+1}$, or equivalently the slope
\be
r_f=-\sum_{a=1}^{f} \frac{l_a}{p_a+1}\, .
\ee
The resultant line charge from this general potential is 
\begin{align}
\lambda(\eta)=\begin{cases}
\eta \, \Big(r_f + \sum_{a=1}^{f} l_a\Big)\, ,& 0\leq \eta\leq n_1\, ,\\
\qquad \qquad\vdots &\qquad \vdots\\
\eta\, \Big(r_f+\sum_{a=i}^{f} l_a\Big)+\sum_{a=1}^{i-1}l_a n_a\, ,& n_{i-1}\leq \eta\leq n_i\, ,\\
\qquad \qquad \vdots &\qquad \vdots\\
\eta\, r_f+\sum_{a=1}^{f} n_a l_a\, ,& n_f\leq \eta\leq n_{f+1}\, ,
\end{cases}\label{eq:genlinecharge}
\end{align}
where
\be
r_{f}=-\frac{\sum_{a=1}^{f} n_{a} l_a}{n_{f+1}}\, .\label{eq:rfslope}
\ee
Generically the slope $r_f$ is not integer, but becomes integer if $\sum_{a=1}^{f} n_{a} l_a= m n_{f+1}$ for some integer $m$. 
The end-point $n_{f+1}$ has been fixed by solving $\lambda(n_{f+1})=0$ and assuming that there is no other positive 0 of the line charge.
If the line charge has a zero at a positive value smaller than $n_{f+1}$ we must cut off the line-charge there. Note by construction that the line charge is convex and has kinks at $n_a$ with change of slope $l_a$.\footnote{If two of the $n_a$ are equal then the change in the slope is the sum of the 2 $l$'s and there exists a value for $p$ such that the two potentials can be described by a single potential.}
The conditions arising from flux quantisation impose (this follows straightforwardly from section \ref{sec:electro} so we do not repeat the analysis)
\begin{align}
\frac{2}{\pi\lp^3}n_a=N_a\in \mathbb{Z}\, ,\quad \frac{2}{\pi\lp^3} n_a l_a\in \mathbb{Z}\, ,\label{eq:genfluxquant}
\end{align}
for all $a$. We may solve all these conditions by defining
\be
N=\frac{2}{\pi \lp^3}\frac{g}{\sum_{a=1}^{f} l_a P_a}\, ,
\ee
where 
\be
P_a=\prod_{b\neq a}(p_b+1)\, ,\quad P=\prod_{a=1}^{f}(p_a+1)\, ,\quad g=\text{gcd}\Big[P \sum_{a=1}^{f} n_a l_a,\sum_{b=1}^{f}l_b P_b\Big]\, .
\ee
We end up with $2 f+1$ independent quanta, the $f$ changes of slope $l_a$ and the $f+1$ positions $N_{a}$, $N_{f+1}$ included. We will relate combinations of these quanta to the field theory parameters. The introduction of the integer $N$ is useful for understanding the holographic limit of the solution, however in comparing to the field theory it is more useful to work with the integers $N_a$ and $N_{f+1}$. 

\begin{figure}[h!]

\begin{subfigure}{0.5\textwidth}
\centering

  \includegraphics[width=0.9\linewidth]{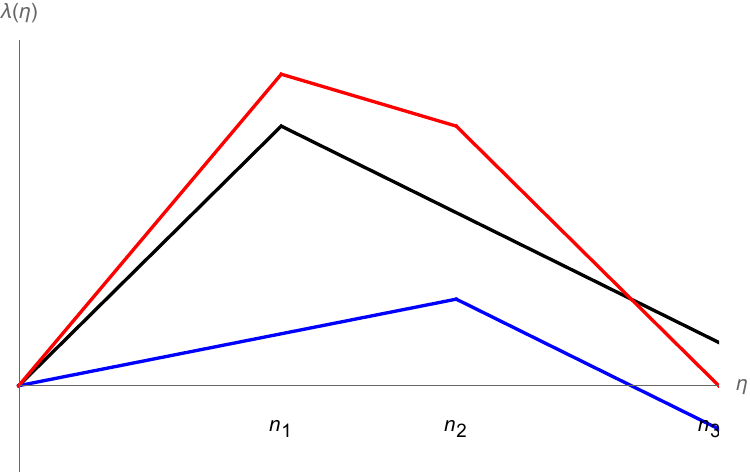}
  \caption{Plot for $p_1=2$, $p_2=\frac{2}{5}$.}
  \end{subfigure}
   \begin{subfigure}{0.5\textwidth}
\centering

  \includegraphics[width=0.9\linewidth]{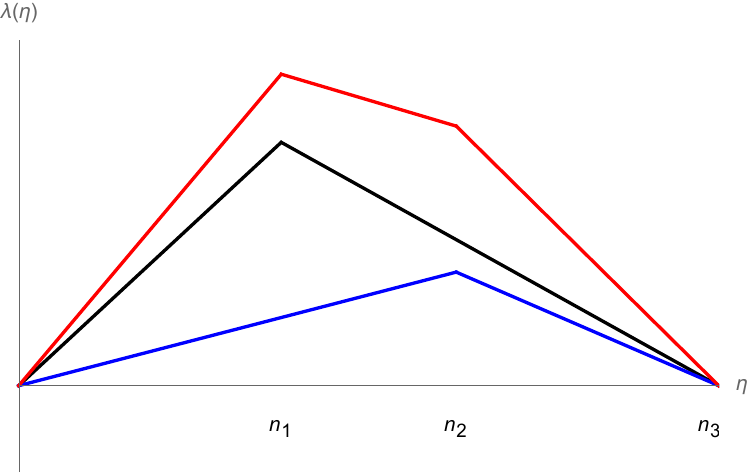}
\caption{Plot for $p_1=\frac{5}{3}$, $p_2=\frac{3}{5}$.}
  \end{subfigure}
  
    \begin{subfigure}{0.5\textwidth}
\centering

  \includegraphics[width=0.9\linewidth]{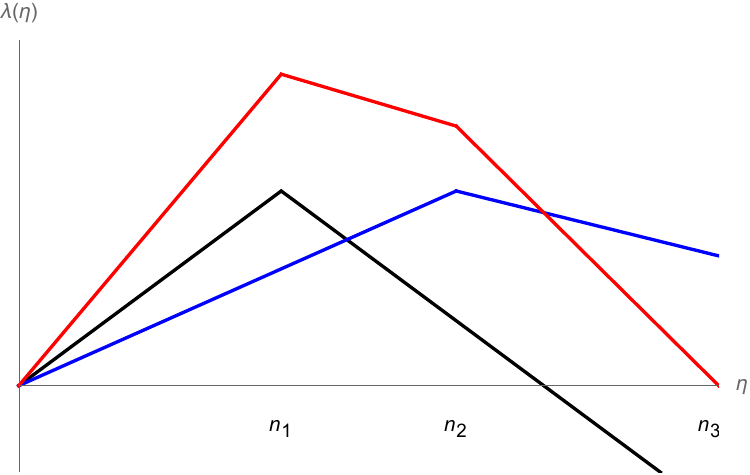}
\caption{Plot for $p_1=1$, $p_2=\frac{9}{5}$.}
  \end{subfigure}
    \begin{subfigure}{0.5\textwidth}
\centering

  \includegraphics[width=0.9\linewidth]{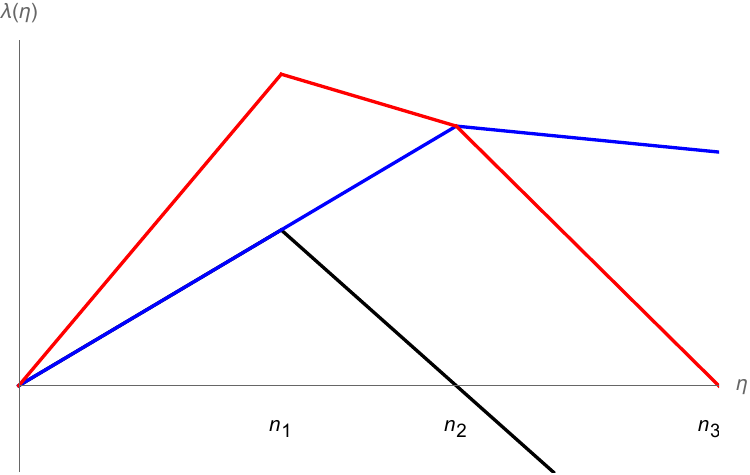}
\caption{Plot for $p_1=\frac{2}{3}$, $p_2=6$.}
  \end{subfigure}
  
    \captionsetup{width=.95\linewidth}
  \caption{We plot various choices of the building blocks giving rise to the same theory. The red line is the line charge of the composite potential which is identical in all four figures with end-point $n_3=8$. The black line is the line charge of the first building block giving the kink at $n_1=3$ with change of slope $l_1=15$. The blue line is the second building block with kink at $n_2=5$ and change of slope $l_2=7$. }
  \label{fig:linechargedegen}
  
\end{figure}

One may wonder whether it is possible to extract out the building blocks used to construct the potential. One can recover information about the location of the monopoles of the building blocks and the associated change of slope, however there is no way to recover the information about $p$. One can construct multiple choices of building block with fixed $n_a$ and $l_a$ but varying $p_a$ giving rise to the same final configuration. In figure \ref{fig:linechargedegen} we present four examples of different building blocks giving rise to the same final theory with two kinks. It would be interesting to understand whether one can understand this as the collision of regular punctures giving rise to the composite regular puncture.  

\begin{figure}[h!]

\centering
 \includegraphics[width=0.9\linewidth]{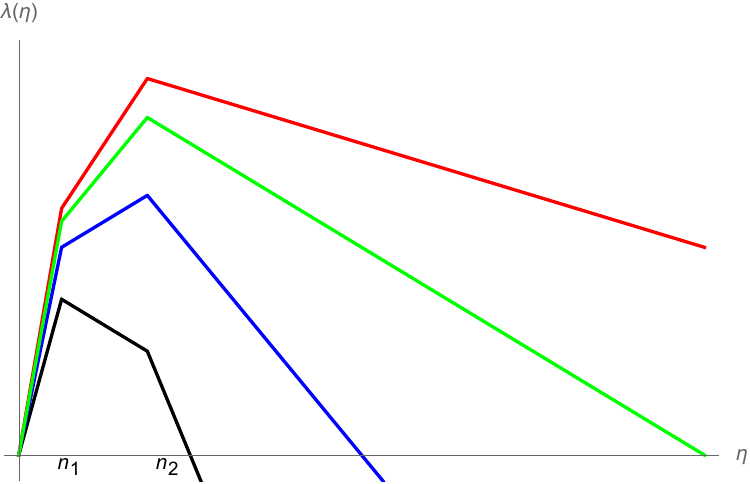}
  
    \captionsetup{width=.95\linewidth}
  \caption{We plot four choices of line-charge with two kinks. Each has identical change of slopes and kink locations but with different end-point $n_{3}$. As we will show later from the kink locations and changes of slope one may construct a Young diagram labelling the choice of regular puncture of the sphere. On the other hand the location $n_{f+1}$ will be data for the irregular puncture. The solutions giving rise to the line charges plotted here are therefore dual to theories with the same regular puncture but with a \emph{different} irregular puncture. }
  \label{fig:linechargesameY}
  
\end{figure}

Another interesting aspect of the superposition of these solutions is that one can find solutions where the kinks and change of slope at the kinks are identical but the location of the zero at $n_{f+1}$ is changed. This hints that the information contained in the kinks labels a regular puncture whilst the location of the non-trivial zero contains irregular puncture data. This will motivate the proposal for the holographic dictionary we present in section \ref{sec:fieldtheory}. We have plotted a few examples of this behaviour in figure \ref{fig:linechargesameY}. 

The domain of $(\rho,\eta)$ is fixed in a similar manner to the single kink case. We restrict to the positive quadrant in the $(\rho,\eta)$-plane. Along $\rho=0$ the circle with coordinate $\phi$ shrinks smoothly, whilst along $\eta=0$, $\dot{V}=0$ and the $S^2$ shrinks smoothly. The final bound is obtained by solving $\dot{V}=0$ away from $\eta=0$ and is the analogue of the ellipse in the single kink case. To understand the shape of the domain it is useful to recall that the level sets of the function $w$ defined in \eqref{eq:wdef} are ellipses, with focus at $n$. We can then consider the intersection of the various level sets of the building blocks $\dot{\mathcal{V}}(\rho,\eta;p_a,l_a,n_a)$. With the level sets defined in \eqref{eq:levelsets}, the condition $\dot{V}=0$ becomes
\be
\sum_{a=1}^{f}\frac{l_a}{p_a+1} \alpha_a=0\, .\label{eq:genlevelseteq}
\ee
We keep fixed $p_a, l_a, n_a$ and solve this for choices of $\alpha_a$ subject to $-1\leq 2\alpha_a\leq p_a$. Clearly for arbitrary values of the $\alpha_a$'s there may not be a solution, however for a fixed set of $\alpha$'s with solution this gives rise to a combination of $f$ curves (either ellipses or hyperbolas) which intersect at a point in the positive quadrant. Plotting all possible points as we vary the possible $\alpha$'s gives rise to a smooth curve going from $(\rho,\eta)=(0,n_{f+1})$ to a point along the $\eta=0$ axis. This point along the $\eta=0$ axis is a solution to a set of $f$ coupled equations and we are unable to give a closed form expression for this point. One must solve
\be
\frac{n_a^2(p_a-2 \alpha_a)\big(p_a-2(\alpha_a+1)\big)}{(1+2 \alpha_a)^2}=\rho_*^2\, ,\quad \forall ~a\in[1,f]\, ,
\ee
subject to \eqref{eq:genlevelseteq} for the $f-1$ remaining $\alpha_a$'s and finally for the location $\rho_*$.

%%%%%%%%%%%%%%%%%%
%%					Observables
%%%%%%%%%%%%%%%%%%

\subsection{Observables}\label{sec:multiobs}

{\bf Central charge}

We can now compute the central charge of the solution using the general form from \eqref{eq:agen}
\begin{align}
a&=\Big(\frac{2}{\pi \lp^3}\Big)^3\frac{1}{4}\int_{0}^{n_{f+1}}\lambda(\eta)^2\dd\eta\nonumber\\
&=\Big(\frac{2}{\pi \lp^3}\Big)^3\frac{1}{12}\sum_{a=0}^{f}\frac{\lambda(n_{a+1})^3-\lambda(n_{a})^3}{r_a }\label{eq:agrav}\\
&=\Big(\frac{2}{\pi \lp^3}\Big)^3\frac{1}{12}\sum_{a=0}^{f}\Big[ r_a^2 (n_{a+1}^3-n_a^3)+3 r_a m_a(n_{a+1}^2-n_{a}^2)+3m_a^2(n_{a+1}-n_{a})\Big]\nonumber\, ,
\end{align}
where we have used the short-hand
\be
m_{a}=\sum_{b=1}^{a} n_{b}l_{b}\, .
\ee
It is useful to split the various terms up, we may write the first term as
\be
\Big(\frac{2}{\pi \lp^3}\Big)^3\sum_{a=0}^{f} r_{a}^2 (n_{a+1}^3-n_{a}^3)=\hat{N}^2 N_{f+1}+\sum_{a=0}^{f-1}\bigg[\Big(\sum_{b=a+1}^{f}l_b\Big)^2 (N_{a+1}^3-N_a^3)\bigg]-2\frac{\hat{N}}{N_{f+1}}\sum_{a=0}^{f} N_{a}^3 l_{a}\, ,
\ee
where
\be
\hat{N}=\sum_{a=1}^{f} N_a l_a\, ,\label{eq:Nhatgravdef}
\ee
and we have used the quantisation conditions \eqref{eq:genfluxquant}. The second term gives
\be
\Big(\frac{2}{\pi \lp^3}\Big)^3\sum_{a=0}^{f} r_a m_a(n_{a+1}^2-n_{a}^2)=-\hat{N}^2 N_{f+1}+\sum_{a=0}^{f-1}\big(N_{a+1}^2-N_a^2\big)\Big[\sum_{b=a+1}^{f}l_b\Big]\sum_{c=1}^{a}l_c N_c+\frac{\hat{N}}{N_{f+1}}\sum_{a=1}^{f}N_a^3 l_a 
\ee
whilst the third term gives
\be
\Big(\frac{2}{\pi \lp^3}\Big)^3\sum_{a=0}^{f} m_a^2(n_{a+1}-n_{a})=N_{f+1} \hat{N}^2-N_{f}\hat{N}^2 +\sum_{a=0}^{f-1}\big[\hat{N}-\sum_{b=a+1}^{f}N_b l_b\big]^2 (N_{a+1}-N_{a})\, .
\ee
Putting everything together we have
\begin{align}
a&=\frac{1}{12 N_{f+1}}\bigg\{\hat{N}^2 N_{f+1}^2+\hat{N}\sum_{a=1}^{f} N_a^3  l_a+ N_{f+1}\sum_{a=0}^{f-1}\bigg((N_{a+1}^3-N_a^3)\Big[\sum_{b=a+1}^{f}l_b\Big]^2\nonumber\\
&+ \Big[\sum_{b=a+1}^{f}l_b\Big]\Big[\sum_{c=1}^{a} N_c l_c\Big]+(N_{a+1}-N_a)\Big[\sum_{b=1}^{a} N_b l_b\Big]^2 \bigg)-N_f N_{f+1}\hat{N}^2\bigg\}\, .\label{eq:asugra}
\end{align}

\vspace{4mm}

\noindent {\bf Scaling dimensions of BPS probe M2-branes}

\noindent Using the result in \eqref{eq:dim1} the scaling dimensions of BPS probe M2-branes located at the kinks is given by
\be
\Delta(\mathcal{O}_a)=\frac{2}{\pi \lp^3}\lambda(n_a)= N_a\Big( \sum_{b=a}^{f}l_b-\frac{\hat{N}}{N_{f+1}}\Big)+\sum_{b=1}^{a-1}N_b l_b\, .
\ee
It is convenient to rewrite this by first defining 
\be
A_a=N_a\sum_{b=a}^{f} l_b+\sum_{b=1}^{a-1} N_b l_b
\ee
then
\begin{align}
\Delta(\mathcal{O}_{a})&=A_a \frac{N_{f+1}-\hat{N}}{N_{f+1}}+(A_a-N_a)\frac{\hat{N}}{N_{f+1}}\\
&=A_a -N_a\frac{\hat{N}}{N_{f+1}}\, .\label{eq:gravconfdim}
\end{align}
We will see that the combination $A_a-N_a$ appears later in the field theory section and can be obtained by studying a Young tableau.\\

\noindent {\bf Flavour symmetries}

\noindent  We may also compute the flavour central charge due to the flavour groups arising at the location of the kinks. From \eqref{eq:flavourgrav} we have that the central charge at the $a$'th kink is
\be
k_{F_a}=\frac{4}{\pi \lp^3} \lambda(n_a)\, ,
\ee
which gives
\be
k_{F_a}=2 \bigg[ N_a\Big( \sum_{b=a}^{f}l_b-\frac{\hat{N}}{N_{f+1}}\Big)+\sum_{b=1}^{a-1}N_b l_b\bigg]\, ,
\ee
with $\hat{N}$ as given in \eqref{eq:Nhatgravdef}. Note that this is twice the scaling dimensions of the BPS operators considered above.

%%%%%%%%%%%%%%%%%%%%%%%%%%%%%%%%%%%%%%%%%%%%
%%							Field Theory
%%%%%%%%%%%%%%%%%%%%%%%%%%%%%%%%%%%%%%%%%%%%

\section{Field theory dual}\label{sec:fieldtheory}

In \cite{Bah:2021hei} the dual field theory of the disc solution that we reviewed in section \ref{sec:single} was identified to be the 4d $\mathcal{N}=2$ theory 
\be
\Big(I_{\hat{N},\hat{k}},\hat{Y}_l\Big)\, ,
\ee
with regular puncture given by the Young diagram $Y_l$ consisting of the rectangle with $l$ columns and $\hat{N}/l$ rows. This class of SCFTs are of Argyres--Douglas type and arise from the low-energy limit of $\hat{N}$ M5-branes wrapped on a sphere with irregular puncture of type $I_{\hat{N},\hat{k}}$ and regular puncture with associated Young diagram $Y_l$. Equivalently, they are the compactification of the 6d $\mathcal{N}=(2,0)$ $A_{\hat{N}-1}$ theory on the same punctured sphere.

Motivated by this we conjecture that the solutions we constructed in section \ref{sec:multi} are the holographic duals of the 4d $\mathcal{N}=2$ theories
\be
\Big(I_{\hat{N},\hat{k}}\, ,Y \Big)\, ,
\ee
where the Young diagram $Y$ is a general partition of $\hat{N}$, not necessarily rectangular.

To keep this paper as self-contained as possible and to clarify the notation we will use, we first review the Argyres--Douglas theories with emphasis on the observables that we can match with the holographic solutions in section \ref{sec:ADreview}, the reader familiar with these theories and their notation may safely skip this section. We then proceed to explain how to read off the holographic dictionary between the gravity solutions in section \ref{sec:multi}, in particular from the data contained in the line charge in equation \eqref{eq:genlinecharge}, and the parameters of the field theory. We show that the leading order contributions to the central charges from gravity match the field theory results and in addition compare the scaling dimensions of certain BPS operators. 

%%%%%%%%%%%%%%%%%%%%%%%%%%%%%%%%%%%%%%%%%%%%
%%%%%%%					AD review
%%%%%%%%%%%%%%%%%%%%%%%%%%%%%%%%%%%%%%%%%%%%

\subsection{Review of Argyres--Douglas theories}\label{sec:ADreview}

Argyres--Douglas theories describe a set of 4d $\mathcal{N}=2$ SCFTs which admit fractional scaling dimensions of the Coulomb branch operators and dimensional coupling constants. They were first found in \cite{Argyres:1995jj,Argyres:1995xn} as a point on the Coulomb branch of pure $\mathcal{N}=2$ SU$(3)$ gauge theory and have been extended to many different constructions since \cite{Eguchi:1996vu,Shapere:1999xr,Cecotti:2010fi,Xie:2012hs,Cecotti:2012jx,Wang:2015mra}. They are intrinsically strongly coupled theories and because of the non-integer Coulomb branch operators the conformal fixed points cannot be described by a $\mathcal{N}=2$ Lagrangian gauge theory.\footnote{Despite this there are 4d $\mathcal{N}=1$ Lagrangian gauge theories that flow to Argyres Douglas theories, see for example \cite{Maruyoshi:2016tqk,Maruyoshi:2016aim,Agarwal:2016pjo,Agarwal:2017roi,Benvenuti:2017bpg}. }

One can engineer Argyres--Douglas theories using geometric engineering, in particular there are constructions in both type IIB and in M-theory. We will be concerned with the class of theories which can be obtained by compactifying the 6d $\mathcal{N}=(2,0)$ theory of $\mathfrak{g}$=ADE type on a punctured sphere. If the punctures of the sphere are all of regular type one obtains theories of class $\mathcal{S}$, with integer scaling dimensions. If instead, one allows for irregular singularities one can obtain Argyres--Douglas theories, \cite{Bonelli:2011aa,Xie:2012hs,Wang:2015mra}.

One can only obtain a 4d $\mathcal{N}=2$ SCFT with an irregular puncture by compactifying on a sphere and not on a higher order genus Riemann surface. The complex coordinate of the Riemann surface should transform non-trivially under the U$(1)_R$ R-symmetry in the presence an irregular puncture. As such, the puncture must be placed at a fixed point of this rotational symmetry. In the space of Riemann surfaces, only the sphere admits a U$(1)$ action with fixed points. It follows that we may place an irregular puncture at one of the poles of the sphere and at most one regular puncture at the other. Any other configurations containing an irregular puncture, whether it be a different Riemann surface or with more than one regular puncture, will not give rise to a $\mathcal{N}=2$ SCFT \cite{Xie:2012hs}.\footnote{Note that if there is no irregular puncture the complex coordinate of the Riemann surface does not transform under the U$(1)_R$ R-symmetry and therefore the above analysis does not apply. One can allow for an arbitrary number of regular punctures for any Riemann surface and obtain a theory of class $\mathcal{S}$.}

The possible irregular punctures were classified in \cite{Xie:2012hs} using the Hitchin equation. Similar to the classification of theories of type $\mathcal{S}$ one can identify the Seiberg--Witten curve with the spectral curve of the Hitchin system on the sphere. The Hitchin system consists of a pair of one-forms, $(A,\phi)$ one a gauge field and the other a Higgs field, each valued in some Lie Algebra $\mathfrak{g}$. Hitchin's equations are equivalent to imposing that the curvature of $\mathcal{A}=A+\ii\phi$ is flat. Punctures correspond to singular solutions to Hitchin's equations, the form of which are constrained. Let $\mathrm{z}$ be a complex coordinate on the sphere and $\Phi(\mathrm{z})$ be the holomorphic part of the Higgs field $\phi$. Consider the six-dimensional $A_{\hat{N}-1}$ $(2,0)$ theory, i.e. the Hitchin system is SU$(\hat{N})$ valued. Then, for an irregular singularity at $\mathrm{z}=0$ the Higgs field near the singularity behaves as one of the following three forms
\begin{align}
\text{Type I}&\, ,\quad \text{I}_{\hat{k},\,\hat{N}}:\quad &\Phi(\mathrm{z})&=\frac{T}{\mathrm{z}^{2+r}}+\cdots\, ,\quad &r=&\frac{\hat{k}}{\hat{N}}>-1\, , \nonumber\\
\text{Type II}&\, ,\quad \text{II}_{\hat{k},\,\hat{N}}: &\Phi(\mathrm{z})&=\frac{T}{\mathrm{z}^{2+r}}+\cdots\, ,\quad &r=&\frac{\hat{k}}{\hat{N}-1}>-1\, ,\\
\text{Type III}&\, ,\quad \text{III}_{Y_l,\cdots,Y_{1}}:&\Phi(\mathrm{z})&=\frac{T_l}{\mathrm{z}^{l}}+\cdots+\frac{T_1}{\mathrm{z}}+\cdots\, ,\quad &Y_{l}\subset&\,  Y_{l-1}\subset\cdots \subset Y_1\, .\nonumber
\end{align}
The final dots denote non-divergent terms and the matrices $T$ are SU$(\hat{N})$ valued diagonal matrices. For type I and type II theories they have $\hat{N}$ independent eigenvalues whilst for type III theories the degeneracy of the eigenvalues is encoded in the corresponding Young tableaux $Y_i$. The inclusion of a regular puncture with one of the above three irregular punctures gives rise to a type IV theory. These theories are labelled by the choice of irregular puncture $P$ and regular puncture $Y$, and will be denoted $(P,Y)$. In this work we will be interested in type IV theories where the irregular puncture is of type I. 
The Seiberg--Witten curve takes the form
\be
0=x^{\hat{N}}+\mathrm{z}^{\hat{k}}+\sum_{(m,n)\in S} v_{m,n}\mathrm{z}^{n} x^{m}\, ,
\ee
where $S$ is defined by the Newton polygon for the theory. The scaling dimension of $x$ and $\mathrm{z}$ are fixed by taking the Seiberg--Witten differential $x\,\dd\mathrm{z}$ to have dimension $1$, and therefore
\be
[x]=\frac{\hat{k}}{\hat{k}+\hat{N}}\, ,\qquad [\mathrm{z}]=\frac{\hat{N}}{\hat{k}+\hat{N}}\, .
\ee
We refer the reader eager for more details on the classification of irregular punctures and Newton polygons, after this short and simplified review, to \cite{Xie:2012hs,Wang:2015mra}.

\subsection{Observables}

{\bf Central charges}

\begin{figure}[h!]
\setlength{\unitlength}{0.14in}
\centering
\begin{tikzpicture}
 %%%%%%%%%% 1st block %%%%%%%%%%%%%%%%%%%%%%%%
 \draw (0,0) rectangle (3,7); 
 \draw[dashed,gray] (2.2,0) -- (2.2,7);
 \draw[dashed,gray] (0.8,0) -- (0.8,7);
 \draw[dashed,gray] (0,0.8) -- (3,0.8);
 \draw[dashed,gray] (0,6.2) -- (3,6.2);
 \draw [<->] (-0.3,0) -- (-0.3,7);
 \draw [<->] (0,7.3) -- (3,7.3);
 \draw (-0.6,3.5) node {$\hat{n}_f$};
 \draw (1.5,7.6) node {$l_f$};
 \draw (0.4,0.4) node {$1$};
 \draw (2.6,0.4) node {$l_f$};
 \draw (1.5,0.4) node {$\ldots$};
 \draw (1.5,6.6) node {$\ldots$};
 \draw (2.6,6.6) node {{\large $\hat{N}$}};
 \draw (2.6,3.5) node {$\vdots$};
 %%%%%%%%%% 2nd block %%%%%%%%%%%%%%%%%%%%%%%%%
 \draw (3,6) -- (5.5,6) -- (5.5,0) -- (3,0);
 \draw[dashed,gray] (4.7,0) -- (4.7,6);
 \draw[dashed,gray] (3,5.2) -- (5.5,5.2);
 \draw (5.11,5.6) node {$i_{f-1}$};
 \draw (4,5.6) node {$\ldots$};
 \draw (5.1,3) node {$\vdots$};
 \draw (6.5,2) node {$\cdots$};
 %%%%%%%%%% ath block %%%%%%%%%%%%%%%%%%%%%%%%%
 \draw (7.5,0) rectangle (9.5,4);
 \draw[dashed,gray] (8.7,0) -- (8.7,4);
 \draw[dashed,gray] (7.5,3.2) -- (9.5,3.2);
 \draw [<->] (9.8,0) -- (9.8,4);
 \draw (10.1,2) node {$\hat{n}_a$};
 \draw [<->] (7.5,4.3) -- (9.5,4.3);
 \draw (8.5,4.6) node {$l_a$};
 \draw (9.1,3.6) node {$i_a$};
 \draw (8.2,3.6) node {$\ldots$};
 \draw (9.1,2) node {$\vdots$};
 \draw (10.9,2) node {$\cdots$};
 %%%%%%%%%% final block %%%%%%%%%%%%%%%%%%%%%%%%%
 \draw (11.7,0) rectangle (14,2.8);
 \draw[dashed,gray] (11.7,2) -- (14,2);
 \draw[dashed,gray] (13.2,0) -- (13.2,2.8);
 \draw (13.6,2.4) node {$i_1$};
 \draw (12.5,2.4) node {$\ldots$};
 \draw (13.6,1.3) node {$\vdots$};
  \draw [<->] (14.3,0) -- (14.3,2.8);
   \draw (14.6,1.4) node {$\hat{n}_1$};
    \draw [<->] (11.7,3.1) -- (14,3.1);
   \draw (12.85,3.4) node {$l_1$};

\end{tikzpicture}
\captionsetup{width=.95\linewidth}
\caption{Our conventions for the Young diagram and its labelling. The boxes are labelled from left to right, bottom to top and give a partition of $\hat{N}$. The corner boxes are distinguished and we denote them by $i_a$ with $a\in\{1,..,f\}$ and are defined in \eqref{eq:iadef}. By construction $\hat{N}= i_f$.}
\label{fig:Ydiagram} 
\end{figure}
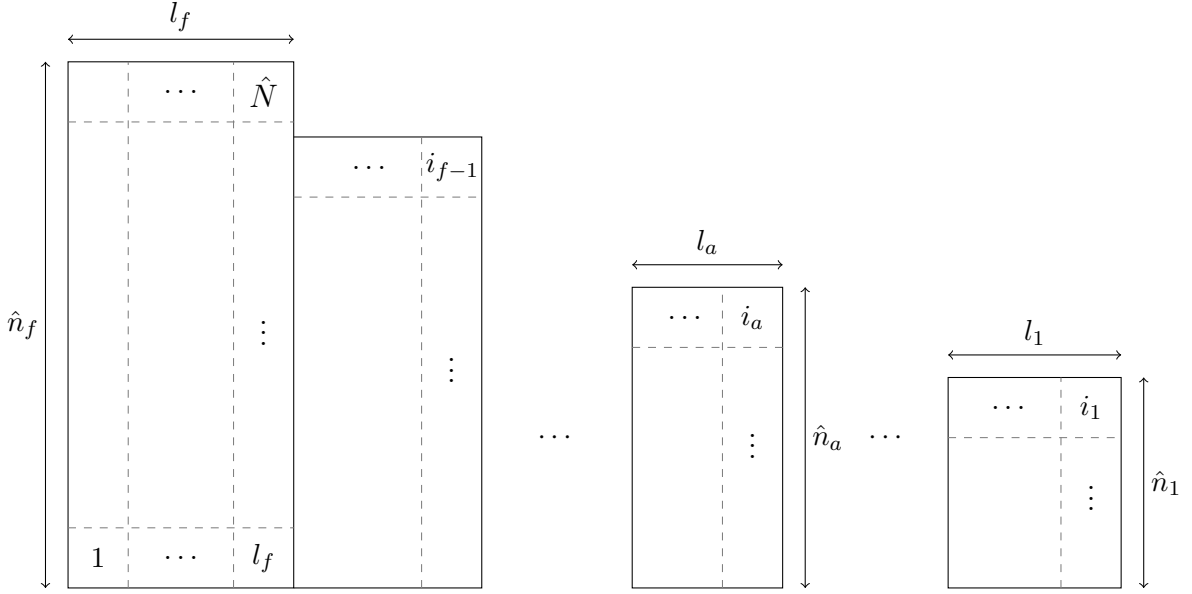

\noindent The main observable that we wish to compare to our gravity solutions is the leading order contribution to the `$a$'-central charge. There is a ``straightforward" way of obtaining the central charge from knowledge of the central charge of the maximal puncture theory and the regular puncture theory \cite{Giacomelli:2020ryy}. We will focus on the leading order terms for large $\hat{N},\hat{k}$ and suppress the subleading terms. As explained in \cite{Giacomelli:2020ryy} the central charges of the $(I_{\hat{N},\hat{k}},Y)$ theory are equal to the sum of four pieces: 
\begin{align}
a&= a_{Y}+ \frac{\hat{N}}{\hat{N}+\hat{k}}\frac{ 6 I_{\rho Y}-\hat{N}(\hat{N}^2-1)}{12}+ a_{I_{\hat{N},\hat{k}}}\,,\label{eq:acentral}\\
c&= c_{Y}+ \frac{\hat{N}}{\hat{N}+\hat{k}}\frac{ 6 I_{\rho Y} - \hat{N}(\hat{N}^2-1)}{12}+ c_{I_{\hat{N},\hat{k}}}\, .\label{eq:ccentral}
\end{align}
Here $a_{Y}$ and $c_{Y}$ are the standard contributions from the puncture $Y$, $I_{\rho Y}$ is the embedding index of SU$(2)$ in SU$(\hat{N})$ associated to the nilpotent vacuum expectation value which is turned on to deform the full puncture to the puncture $Y$. Finally, $a_{I_{\hat{N},\hat{k}}}$ and $c_{I_{\hat{N},\hat{k}}}$ are the central charges of the $I_{\hat{N},\hat{k}}$ theory, i.e. the theory with just the irregular puncture. Keeping only the leading order terms the last terms contribute
\be
a_{I_{\hat{N},\hat{k}}}=c_{I_{\hat{N},\hat{k}}}= \frac{\hat{k}^2\hat{N}^2}{12(\hat{k}+\hat{N})}\, ,
\ee
to the total central charge. To understand the contributions from the regular puncture let us set the conventions for the Young diagram. We will view the Young diagram as a series of rectangles with height $\hat{n}_a,\,  a\in (1,f)$ and width $l_a$. In total the Young diagram consists of $\hat{N}=\sum_{a=1}^{f} \hat{n}_al_a$ boxes, thus giving a partition of $\hat{N}$, see figure \ref{fig:Ydiagram}.

The embedding index in these conventions is 
\be
I_{\rho Y}=\frac{1}{6} \sum_{a=1}^{f} \hat{n}_a(\hat{n}_{a}^2-1)l_{a}\, ,
\ee
and this regular puncture leads to the flavour symmetry 
\be
G_{\text{flavour}}=\text{S}\bigg( \prod_{a=1}^{f} \text{U}(l_a)\bigg)\, .
\ee
The final, as of yet unspecified, contributions are from $a_{Y}$ and $c_{Y}$. Following the conventions in \cite{Bah:2018gwc} we have (dropping some obviously subleading terms)
\begin{align}
a_{Y}&=\frac{1}{6} \hat{N}^3+\frac{1}{24}\big(n_h(Y)+5 n_v(Y)\big)\, ,\quad c_{Y}=\frac{1}{6}\hat{N}^3+\frac{1}{12}\big(n_h(Y)+2 n_v(Y)\big)\, ,
\end{align}
where $n_{h}(Y)$ and $n_{v}(Y)$ are the effective number of hypermultiplets and vector multiplets from the regular puncture. We have
\begin{align}
n_v(Y)&=-\frac{1}{2}\big(\hat{N}^2-1\big)+\sum_{a=1}^{f}\sum_{i= \hat{n}_{a-1}+1}^{\hat{n}_{a}}\bigg( \sum_{b=1}^{a-1}\Big[(\hat{n}_b-\hat{n}_{b-1})\sum_{c=b}^{f}l_c\Big]+(i-\hat{n}_{a-1})\sum_{b=a}^{f}\l_b\bigg)^2 \label{eq:nv}\\
n_{h}(Y)&=n_v(Y)-\frac{1}{2}+\frac{1}{2}\sum_{a=1}^{f} \sum_{b=1}^{a}\sum_{c=b}^{f} l_a (\hat{n}_b-\hat{n}_{b-1})l_{c}\, .\label{eq:nh}
\end{align}
Note that in the holographic limit $n_h(Y)=n_v(Y)$ since the last two terms of \eqref{eq:nh} are subleading. Moreover, the first terms in \eqref{eq:nv} are also subleading and can therefore be dropped too leaving only the summation term. Note in addition that the summation term contains terms with different scaling properties in the holographic limit. \\

\noindent {\bf BPS operator scaling dimensions}

\noindent There are a set of distinguished BPS operators in the spectrum that we can compute the dimensions of and match to the supergravity solution. To proceed we label the boxes of the Young diagram, working from left to right and bottom to top. To the $j$'th box we associate the number $s_j=j- \text{height}(j)$. The $f$ boxes on the far right which have no box above them are distinguished and we denote them by $i_a$. They correspond to operators associated to monomials of the form
\be
\frac{\mathcal{O}_{a}}{\mathrm{z}^{s_{i_a}}}x^{\hat{N}-{i_a}}\, ,
\ee
in the Seiberg--Witten curve.
We may compute their dimension by using the dimension of the coordinates $x,\mathrm{z}$:
\be
[x]=\frac{\hat{k}}{\hat{k}+\hat{N}}\, ,\quad [\mathrm{z}]=\frac{\hat{N}}{\hat{k}+\hat{N}}\, .
\ee
It thus follows that the operator $\mathcal{O}_{a}$ has dimension
\begin{align}
[\mathcal{O}_a]&=\big(\hat{N}-(\hat{N}-i_a)\big)[x]+s_{i_a} [z]\nonumber\\
&=i_a - \text{height}({i_a}) \frac{\hat{N}}{\hat{k}+\hat{N}}\, .
\end{align}
For the Young diagram pictured in figure \ref{fig:Ydiagram} and focussing on the $a$'th distinguished box at the $a$'th right-most corner (working form below) we can write
\be
i_a =n_a \sum_{b=a}^{f} l_b+ \sum_{b=1}^{a-1} n_b l_b\, ,\quad \text{height}(i_a)=n_a\, ,\label{eq:iadef}
\ee
and the conformal dimension becomes
\be
\Delta(\mathcal{O}_a)=n_a \sum_{b=a}^{f} l_b+ \sum_{b=1}^{a-1} n_b l_b- n_a \frac{\hat{N}}{\hat{k}+\hat{N}}\, .\label{eq:dimOft}
\ee
\\

\noindent {\bf Flavour central charges}

\noindent With the above operator dimensions in hand it is a simple matter to compute the central charge of the $a$'th non-abelian gauge factor. 
Following the conjecture in \cite{Xie:2013jc} the flavour central charge is twice the conformal dimension of the operator located the corresponding right-most box of the Young diagram. Using the results above we have 
\be
k_{F_a}=2 \bigg[n_a \sum_{b=a}^{f} l_b+ \sum_{b=1}^{a-1} n_b l_b- n_a \frac{\hat{N}}{\hat{k}+\hat{N}}\bigg]\, .
\ee

%%%%%%%%%%%%%%%%%%%%%%%%%%%%%%%%%%%%%%%%%%%%
%%%%%%%					Holographic dictionary
%%%%%%%%%%%%%%%%%%%%%%%%%%%%%%%%%%%%%%%%%%%%
\vspace{3mm}

\subsection{Setting up the holographic dictionary}\label{sec:dictionary}

We now want to understand how to map the gravity parameters to the field theory parameters introduced in the previous section.
One may ask whether we can use the rules of \cite{Gaiotto:2009gz} to construct a dual quiver theory. The short answer is that generically we cannot apply the rules to construct a quiver. Given that the $(I_{\hat{N},\hat{k}},Y)$ theories are generically non-Lagrangian this is somewhat reassuring since we wish to identify them as a dual pair. 
To see why application of the rules in \cite{Gaiotto:2009gz} for constructing a quiver fails one should note the non-integer slopes of the line-charge for the solutions discussed here. In \cite{Gaiotto:2009gz} they are taken to be integer, which is a requirement for constructing a dual quiver since the values of the line charge at the integer values of $\eta$ become the ranks of the gauge groups in the quiver. We emphasise that having integer slopes for the line charge is not a requirement of a well-defined supergravity solution, only that the change of the slopes at the kinks is integer. For certain choices of the parameters we can make the slope $r_f$ integer, and therefore all the other slopes too, which allows for the construction of a dual quiver theory, however this is not a generic choice one can make and so we will not focus on this possibility.

Recall that the line charge depends on $2f+1$ parameters, the $f$ changes of slopes $l_a$, the $f$ kink locations $n_{a}$ and the end-point $n_{f+1}$. The field theory we conjecture is dual also has $2f+1$ parameters: the integer $\hat{k}$ and the $2 f$ parameters describing the Young diagram giving a partition of $\hat{N}$. We conjecture that the dictionary between the field theory and gravity solutions is
\be
\hat{k}= N_{f+1}-\hat{N}\, ,\quad \hat{N}=\sum_{a=1}^{f} N_{a} l_{a}\, ,\label{eq:dic1}
\ee 
with the identifications
\be
\hat{n}_a=N_a\, ,\quad l_a^{\text{SCFT}}=l_{a}^{\text{SUGRA}}\, ,\label{eq:dic2}
\ee
where the $N_a$ are defined in equation \eqref{eq:genfluxquant} as
\be
\frac{2}{\pi \ls^3}n_a=N_a\in \mathbb{Z}\, .
\ee
Note that this reduces to the identifications made for the disc solution in \cite{Bah:2021hei} as it should.

%%%%%%%%%%%%%%%%%%%%%%%%%%%%%%%%%%%%%%%%%%%%
%%%%%%%					Observables matching
%%%%%%%%%%%%%%%%%%%%%%%%%%%%%%%%%%%%%%%%%%%%

\subsection{Matching of observables}

We now want to match the central charge of the proposed field theory with the gravity result in \eqref{eq:agrav} using the above dictionary. The large $N$ limit is obtained by taking the $\hat{n}_a$'s, $\hat{N}$ and $\hat{k}$ all large and of the same order. It is useful to note that the central charge coming from field theory may be written as
\be
a= \frac{1}{12N_{f+1}}\bigg[N_{f+1}^2 \hat{N}^2+3 N_{f+1}n_{h}(Y)+\hat{N}\sum_{a=1}^{f} \hat{n}_{a}^3l_a \bigg]\, 
\ee
where we have used our dictionary which gives $\hat{k}+\hat{N}=N_{f+1}$. It is helpful to compare the $N_{f+1}$ coefficients. The first and last terms match on the nose with the gravity expressions. The second term is not as obvious but after a short but not particularly enlightening calculation, and removing subleading terms one finds that this term also matches between the field theory and gravity result. We conclude that the central charge of the gravity solution we discuss in section \ref{sec:multi} and the central charge of the $(I_{\hat{N},\hat{k}},Y)$ theories match to leading order given the holographic dictionary in section \ref{sec:dictionary}.

Similarly we may check the match for the scaling dimension of the probe BPS M2-branes we studied in section \ref{sec:multiobs}. These BPS M2-branes were located at the monopole positions or kinks in the line charge. In terms of the Young diagram these should correspond to the distinguished operators at the right-most corners. 
Comparing \eqref{eq:dimOft} with the gravity result in \eqref{eq:gravconfdim} by using the dictionary \eqref{eq:dic1} and \eqref{eq:dic2} we find perfect agreement. Additionally since the flavour central charges on both sides is given by twice the dimensions of the corresponding state/operator they clearly match given the matching of the conformal dimensions.

One interesting point to consider is when one can draw a dual quiver using the prescription in \cite{Gaiotto:2009gz}. As we explained earlier, their prescription for drawing a dual quiver given a line charge, relies on the slope of the line charge being integer. In our setup this follows if the slope $r_f$ in \eqref{eq:rfslope} is integer, which imposes
\be
-r_f=\frac{\hat{N}}{\hat{k}+\hat{N}}=[\mathrm{z}]\in \mathbb{Z}\, .
\ee
Thus, for these theories the scaling dimensions of the operators we consider are integer. The non-integer slope of the line charge is therefore essential for obtaining theories containing operators with non-integer scaling dimensions. 
A final check of the proposed duality is to study the quiver following from the prescription in \cite{Gaiotto:2009gz} when we take the slope of the line charge to be integer. Constructing the quiver following the prescription in \cite{Gaiotto:2009gz}  one can show that the central charges match exactly with the ones presented in \eqref{eq:acentral}-\eqref{eq:ccentral}, including subleading terms. This gives added evidence that our conjectured duality is correct.

%%%%%%%%%%%%%%%%%%%%%%%%%%%%%%%%%%%%%%%%%%%%
%%%%%%%					Additional topics
%%%%%%%%%%%%%%%%%%%%%%%%%%%%%%%%%%%%%%%%%%%%

%%%%%%%%%%%%%%%%%%%%%%%%%%%%%%%%%%%%%%%%%%%%
%%%%%%%					Conclusion
%%%%%%%%%%%%%%%%%%%%%%%%%%%%%%%%%%%%%%%%%%%%

\section{Conclusion}

In this paper we have studied the holographic duals of the Argyres--Douglas theories
\be
\Big(I_{\hat{N},\hat{k}}\, ,Y \Big)\, ,
\ee
with $Y$ an arbitrary regular puncture generalising the work in \cite{Bah:2021hei}. We have shown that there is a perfect match between the field theory and gravity solutions in the holographic limit. 
It would be interesting to go beyond the leading order analysis conducted here and compute subleading corrections on the gravity side, for example to study the gravitational anomaly $a-c$ and check the matching of the central charges more generally. Recently, 4d $\mathcal{N}=2$ SCFTs with $a=c$ have been constructed in \cite{Kang:2021lic} with the holographic duals currently unknown. It is natural to wonder whether using similar singular gravity solutions as those in this paper one can construct the duals of these theories too. 

One of the results of our analysis which deviates from folklore are the necessary constraints a line charge must satisfy to give rise to a well-defined $\mathcal{N}=2$ AdS$_5$ solution using \cite{Gaiotto:2009gz}. We have seen that it is necessary for the integer condition of the slope to be relaxed if the line charge does not plateau. This can be reinterpreted as the $\mathbb{T}^2$ fibration of the internal manifold being quasi-regular as opposed to regular when the slopes are all integer. This leads to M5-branes wrapping cycles in this torus and in the quasi-regular case the dual SCFT is non-Lagrangian and dual operators have fractional dimensions. The gravity analysis then gives a concrete condition on when one can construct a dual quiver and additionally by using the rules in \cite{Gaiotto:2009gz} how to do this. It seems worthwhile to see whether relaxing this condition in previously studied solutions in the literature gives rise to interesting solutions which have been missed previously.

There remain a number of interesting avenues to pursue. For example it remains to construct the holographic duals of spheres with the other types of irregular puncture. We have constructed the holographic dual for the most general type IV theory using a type I irregular puncture however a holographic dual for the theories using the other irregular punctures is currently missing. In a similar vein it would be interesting to construct the holographic duals of the Argyres--Douglas theories constructed from the 6d $\mathcal{N}=(2,0)$ D-series theory, one may obtain some inspiration from \cite{Nishinaka:2012vi}. Further constructing the SCFT duals for the other disc compactifications is an outstanding problem. For the D3 discs, \cite{Couzens:2021tnv,Suh:2021ifj} it is tempting to conjecture that this is dual to $\mathcal{N}=4$ SYM on a punctured sphere which preserves $\mathcal{N}=(2,2)$ in 2d. Similarly the field theory dual for M2-branes on a disc \cite{Suh:2021hef,Couzens:2021rlk} should correspond to ABJM on a punctured sphere, and for the D4-D8 system \cite{Suh:2021aik} to 5d $\mathcal{N}=1$ USp$(2N)$ theory on a punctured sphere. 
In a similar spirit identifying the dual field theories for compactifications on a spindle still remains an open problem.

\section*{Acknowledgments}

We would like to thank Jaewon Song, Monica Jinwoo Kang and Sakura Sch\"afer-Nameki for useful discussions. 
This work was supported by the National Research Foundation of Korea (NRF) grant 
2019R1A2C2004880(CC, HK, NK, YL), 2020R1A2C1008497(CC, HK) and 2018R1A2A3074631(YL).

%%%%%%%%%%%%%%%%%%%%%%
\appendix

%%%%%%%%%%%%%%

\section{Special limits of the 7d metric}\label{app:limits}

In this section we will show that both the Maldacena--Nunez solution \cite{Maldacena:2000mw} and AdS$_7$ can be obtained from the metric in \eqref{eq:7dM5sol} as special limits. We will first study the AdS$_7$ solution before studying the limit for the Maldacena--Nunez solution. We will focus on the uplifted 11d solution \eqref{eq:M511dmetricsimp} in the following for ease of comparison.

\subsection{AdS$_7$}

Since the AdS$_7$ solution should not be supported by any gauge fields we must set them to be pure gauge. We may achieve this by setting $s_1=s_2=0$. Note that this implies that the scalars are set to 1, which is indeed the correct supersymmetric constant values. Changing coordinates to $w=\cosh^2\zeta$, $z=2 \hat{z}$ and $\mu=\sin\theta$ the metric becomes
\be
\dd s^2_{11}=4 \Big(\cosh^2\zeta \dd s^2(\text{AdS}_5)+\sinh^2\zeta \dd \hat{z}^2+\dd\zeta^2\Big)+\dd\theta^2+\cos^2\theta\dd s^2(S^2)+\sin^2\theta \dd\phi_1^2\, ,
\ee
which is the metric on AdS$_7\times S^4$. 
We therefore see that the metric in \eqref{eq:M511dmetricsimp} has as special limit AdS$_7\times S^4$. Note that the $s_1=s_2=0$ is one of the special end-points for the range of well-defined values for $s_1$ at fixed $s_2=0$. It is the value for which $P(w)$ obtains a 4-fold root and $f(w)$ obtains a triple root. 

The potential giving the AdS$_7$ solution in the electrostatic description is
\begin{equation}
V = \eta \log{r} + \frac{1}{2} \cos\theta + \frac{1}{4} \log \left(\frac{1-\cos\theta}{1+\cos\theta}\right)\, ,
\end{equation}
where $\eta = \frac{1}{2}\cos{\theta}\cosh{2\zeta}$, $r = \sin{\theta}\sinh^2{\zeta}$. This can be obtained by taking a suitable limit of the general potential we give in \eqref{eq:potential1}.

\subsection{Maldacena--Nunez solution}\label{app:MNlimit}

We noted in the main text that there is a second special point in $s_1$ parameter space with $s_2=0$ fixed, namely $s_1=-\frac{1}{4}$. At this point the roots of $f(w)$ are $w=0$ twice and $w=\tfrac{1}{2}$ twice. We noted earlier that since $w=\tfrac{1}{2}$ is now a double root the metric at this end-point looks locally like $\mathbb{H}^2$ rather than $\mathbb{R}^2/\mathbb{Z}_l$ as at a generic value $-\tfrac{1}{4}<s_1<0$. Indeed, as we will show momentarily by taking a suitable scaling limit to this point in parameter space we find the Maldacena--Nunez solution.
 
 First, let us take $\mu=\sin\theta$ and perform the rescalings\footnote{The coordinate names have been chosen to match the conventions in \cite{Maldacena:2000mw} and $y$ has no relation to the $y$ introduced earlier in the main text. }
\be
l_1=-\frac{1}{4}\, ,\quad w=\frac{\lambda}{y} +\frac{1}{2}\, ,\quad z= -\frac{x}{\lambda}\, ,\label{eq:MNlimit}
\ee
whilst also performing the coordinate shift
\be
\phi_1\rightarrow \phi_1-\frac{x}{\lambda} \, ,
\ee
which removes a singular term from the gauge field. Expanding around $\lambda=0$ the metric becomes
\begin{align}
\dd s^2_{11}&=\frac{1}{2}(1+\cos^2\theta)^{1/3}\bigg[ 4\dd s^2(\text{AdS}_5)+2\dd\theta^2+\frac{2\cos^2\theta}{1+\cos^2\theta}\dd s^2(S^2)+\frac{4\sin^2\theta}{1+\cos^2\theta}(\dd\phi_1+ y^{-1}\dd x)^2\nonumber\\
&+\frac{2}{y^2}\Big(\dd y^2+\dd x^2\Big)\bigg]\, ,
\end{align}
which is precisely the metric in \cite{Maldacena:2000mw}, with the coordinates used there. We therefore find that the same metric can be globally completed to obtain three seemingly distinct solutions, disc solutions, Maldacena--Nunez solution and pure AdS$_7$. It is interesting to note that the Maldacena--Nunez solution preserves supersymmetry via a topological twist whilst the disc solution does not involve a topological twist but an altogether different mechanism, in particular the Killing spinors of the disc are not independent of the disc coordinates as they would be for a topological twist. 

The electrostatic potential for the Maldacena--Nunez solution is
\begin{equation}
V = \eta \log{r}+ \frac{1}{2} \cos\theta + \frac{1}{4} \log \left(\frac{1-\cos\theta}{1+\cos\theta}\right)\, ,
\end{equation}
where $\eta = \frac{1 + y^2}{4 y} \cos{\theta}$, $r = \frac{y-1}{y+1}$. This can also be obtained from the general potential we provide in \eqref{eq:potential1} after performing the limit in \eqref{eq:MNlimit}.

%%%%%%%%%%%%%%%%%%%%%%%%%%%%%%%%%

%%~~~~~~~~~~~~~~~~~~~~~~~~~~~~~~~~~~~~~~~~~~~~~~~~~~~~~~~~~~~~~~~~~~
%%							LLM for AdS7 and MN
%%~~~~~~~~~~~~~~~~~~~~~~~~~~~~~~~~~~~~~~~~~~~~~~~~~~~~~~~~~~~~~~~~~~

\section{Anomaly inflow}\label{app:anom}

In this section we will study the global symmetries and `t Hooft anomalies of the dual SCFT by using anomaly inflow methods \cite{Freed:1998tg,Harvey:1998bx,Bah:2019rgq}. This section is an extension of the computations performed in \cite{Bah:2021hei} to account for the more general flavour symmetry that our solutions exhibit, as such we will present the bare bones computation when there is no risk of confusion.\footnote{We will use a different presentation in terms of coordinates adapted to the electrostatic description in which our solution is naturally written. } The ultimate goal of this section is to give an independent derivation of the observables studied in the gravity theory and to understand the breaking of a $\mathfrak{u}(1)$ isometry algebra by a St\"uckelberg mechanism as observed in \cite{Bah:2021hei}. 

We want to understand the anomalies of the 4d SCFT living on a stack of M5-branes compactified on a punctured sphere. The 6d internal space, $M_6$ arising in the holographic solution encodes the information about the global symmetries and anomalies of the theory. Following \cite{Bah:2019rgq} one excises a small tubular neighbourhood around the stack of M5-branes, giving the 11d space spacetime a boundary $\partial M_{11}=M_{10}$. $M_{10}$ is a fibration of $M_6$ over the 4d worldvolume $W_4$ on which the 4d SCFT lives. The fibration is determined by the gauge connections for the continuous symmetries of $M_6$ which descend to symmetries of the 4d SCFT. From the magnetic source $G_4$ for the M5-brane stack one can define a closed, globally well-defined four-form $E_4$ on $M_{10}$, with integral periods, which is invariant under the structure group of the fibration and when restricted to $M_6$ reduces to $G_4$.
The presence of the boundary in 11d leads to the topological terms of 11d supergravity no longer being invariant under gauge transformations of background fields on $W_4$. The variation of the Chern--Simons terms give rise to a 10-form linear in the gauge parameters which, via the decent procedure is related to an anomaly 12-form $\mathcal{I}_{12}$ which encodes the anomalies of the theory and takes the form
\be
\mathcal{I}_{12}=-\frac{1}{(2\pi \lp)^9}\frac{1}{3!} E_4\wedge E_4\wedge E_4\, .
\ee
The 6d anomaly polynomial of the 4d SCFT, in the holographic limit, is given by
\be
\mathcal{I}^{SCFT}_{6}= -\int_{M_6}\mathcal{I}_{12}\, ,\label{eq:I6def}
\ee
with the integration over the $M_6$ fibers.

\subsection{Constructing $E_4$}

We now want to construct $E_4$ for the background in the main text. We will gauge all the symmetries in $M_6$. As in \cite{Bah:2021hei} one finds that there is a spontaneous breaking of the continuous symmetries of $M_6$ by a St\"uckelberg mechanism. We will review this quickly for completeness, but refer the reader there for further details. 
The four-form takes the form
\be
G_4=\dd \vol(S^2)\wedge \Big[ \dd f_{\chi}(\rho,\eta)\wedge \dd\chi+\dd f_{\beta}(\rho,\eta)\wedge \dd\beta\Big]\, ,
\ee
with 
\be
f_{\chi}(\rho,\eta)=-\frac{4 \dot{V}^2V''}{\tilde{\Delta}}\, ,\qquad f_{\beta}(\rho,\eta)=2\Big(\frac{\dot{V}\dot{V}'}{\tilde{\Delta}}-\eta\Big)\, .\label{eq:fIs}
\ee
We may gauge the continuous isometries of $M_6$, in total we have the SU$(2)$ and two U$(1)$'s to gauge. This is implemented by the replacements
\be
\dd \beta\rightarrow D\beta=\dd\beta+A^{1}\, ,\quad \dd\chi\rightarrow D\chi=\dd\chi+A^{2}\, ,\quad \dd\vol(S^2)\rightarrow 4\pi e_2\, ,
\ee
with $e_2$ the global angular form of SO$(3)$. Let us introduce the index notation, $I=\{1,2\}\equiv \{\beta,\chi\}$ then the most general form for $E_4$ is
\be
(2\pi\lp)^3 E_4=G_4^{\dd\rightarrow D}+\sum_{I=1}^{2} F^{I}\wedge \omega_I^{\dd\rightarrow D}+\sum_{I,J=1}^{2}\sigma_{(IJ)}F^{I}\wedge F^{J}\, ,
\ee
with $F^{I}=\dd A^{I}$, $\omega_{I}$ two two-forms on $M_6$ and $\sigma_{(IJ)}$ three real scalars on $M_6$. Imposing that $E_4$ is closed allows us to fix the two-forms $\omega_I$ and scalars $\sigma_{(IJ)}$:
\be
\partial_{I} \lrcorner G_4+\dd \omega_I=0\, ,\quad \partial_{(I}\lrcorner \omega_{J)}+\dd \sigma_{(IJ)}=0\, .\label{eq:E4closedcond}
\ee
We have
\be
\partial_{I} \lrcorner G_4=-\dd\Big[4\pi f_{I}(\rho,\eta) e_2 \Big]\, ,
\ee
which are both globally well-defined forms on $M_6$. However, only for $I=2=\chi$ is this an exact form as is necessary to satisfy \eqref{eq:E4closedcond}. To see this consider the degeneration along the smeared brane where the $S^2$ shrinks. For the two-form $f_I(\rho,\eta)e_2$ to be globally well-defined we require that $f_{I}(\rho,\eta)$ vanishes there since the two-sphere shrinks there. From \eqref{eq:fIs} we see that on the smeared brane locus where $\dot{V}=0$ only $f_{\chi}$ vanishes there whilst $f_{\beta}$ does not. The smeared brane source giving rise to the irregular puncture leads to the failure for $f_{\beta}$ to vanish and thus this acts as an obstruction to constructing $E_4$ as above. As explained in \cite{Bah:2021hei} this requires the introduction of an axion which leads to the spontaneous symmetry breaking of the gauge field $A^{\beta}$.

Following \cite{Bah:2021hei} one introduces an axion $\alpha$ with field strength $\hat{f}$ which satisfies
\be
\dd\hat{f}=\sum_{I=1}^{2} q_{I} F^{I}\, ,
\ee
and modifies the ansatz for $E_4$ to take the form
\be
(2\pi\lp)^3E_4=G_{4}^{\dd\rightarrow D}+\sum_{I=1}^{2}F^I\wedge \omega_I^{\dd\rightarrow D}+\sum_{I,J=1}^{2}\sigma_{(IJ)}F^{I}\wedge F^{J}+\hat{f}\wedge \Lambda\, ,
\ee
with 
\be
\Lambda=-4\pi \dd f_{\beta}\wedge e_2\, .
\ee
Closure of $E_4$ implies
\be
\partial_{I}\lrcorner G_4+\dd\omega_I+q_I \Lambda =0\, ,\quad \partial_{(I}\lrcorner \omega_{J)}+\dd\sigma_{(IJ)}=0\, .
\ee
It then follows that imposing closure fixes
\be
\omega_{1}=0\, ,\quad \omega_{2}=-4\pi f_{\chi} e_2\, ,\quad \sigma_{(IJ)}=0\, ,\quad q_{1}=-1\, ,\quad q_2=0\,.
\ee
The Bianchi identity for the axion field strength is
\be
\dd\hat{f}=-F^{1}\, ,\qquad \Rightarrow\qquad \hat{f}=\dd \alpha-A^{1}\, .
\ee
As shown in \cite{Bah:2021hei} this leads to the U$(1)$ gauge field $A^{1}$ becoming massive via a St\"uckelberg mechanism leading to the symmetry being spontaneously broken.

We now want to include the contributions of the flavour symmetries. Recall that at the locations of the kinks of the line-charge the metric is locally the orbifold $\mathbb{R}^4/\mathbb{Z}_{l_a}$ and leads to an SU$(l_a)$ flavour symmetry. The orbifold leads to $l_a-1$ resolution two-cycles on which we may wrap the three-form potential $C_3$ leading to $l_{a}-1$ abelian gauge fields for each kink. Let the resolution two-cycles be denoted by $\omega_{a,i}$ with $i\in \{1,..,l_a-1\}$ and $a\in \{1,..,f\}$. These should be understood to be localised at the kink locations and therefore the intersection of any two of these two-cycles located at different kinks is zero. The intersection pairing then gives
\be
\int_{\mathbb{R}^4/\mathbb{Z}_{l_a}}\omega_{a,i}\wedge \omega_{a,j}=-C_{ij}^{\mathfrak{su}(l_a)}\, .
\ee
We can now turn on background gauge fields $\mathcal{A}_{a,i}$, with field strength $\mathcal{F}_{a,i}$ again with $i\in \{1,..,l_a-1\}$ and $a\in \{1,..,f\}$ by including the term
\be
 \Delta E_4=\sum_{a=1}^{f}\sum_{i=1}^{l_{a}-1}\omega_{a,i}\wedge \mathcal{F}_{a,i}\, .
\ee

\subsection{Anomaly polynomial from anomaly inflow}

We can now insert the ansatz for $E_4$ into the 12-form anomaly polynomial and integrate over $M_6$ to obtain the anomaly polynomial for the 4d theory. To proceed we need a few results that may be extracted from the literature. The Bott--Cattaneo formula \cite{10.4310/jdg/1214460608}
gives
\be
\int_{S^2}e_2=1\, ,\quad \int_{S^2} e_2\wedge e_2=0\, ,\quad \int_{S^2}e_2\wedge e_2\wedge e_2=- c_2(\text{SU}(2)_R)\, ,
\ee
whilst the gauge field $F^{\chi}$ has bundle
\be
F^{\chi}=-4\pi c_1(\text{U}(1)_r)\, ,
\ee
and
\be
\sum_{i,j=1}^{l_a-1}C_{ij}^{\mathfrak{su}(l_a)} \mathcal{F}_{a,i}\wedge \mathcal{F}_{a,j}= 2 c_2(\text{SU}(l_a))\, .
\ee
The final result is due to non-perturbative M2-brane states which enhance the U$(1)^{l_a-1}$ symmetry to the full non-abelian symmetry SU$(l_a)$ as opposed to its Cartan. 

Plugging all these ingredients into \eqref{eq:I6def} we find
\begin{align}
I_{6}^{\text{SCFT}}&=- c_1(\text{U}(1)_r)\wedge c_2(\text{SU}(2)_R) \frac{1}{2(\pi \lp)^3}\int\dd f_{\chi}^2\wedge \dd f_{\beta}+\frac{4}{\pi \lp^3} \sum_{a=1}^{f} \lambda(n_a)c_{1}(\text{U}(1)_r)\wedge c_2(\text{SU}(l_a))\, .
\end{align}
This should be compared with the anomaly polynomial of class $\mathcal{S}$ which reads
\begin{align}
I_{6}^{\text{class} \mathcal{S}}&=(n_v-n_h)\bigg[ \frac{1}{3}c_{1}(\text{U}(1)_r)^3-\frac{1}{12}c_{1}(\text{U}(1)_r)\wedge p_{1}(T^4)\bigg]- n_v c_1(\text{U}(1)_r)\wedge c_2(\text{SU}(2)_R)\nonumber\\
&+\sum_{\text{flavours}} k_F c_1(\text{U}(1)_r)\wedge c_2(\text{SU}(f_a))\, .
\end{align}
We can immediately read off the flavour symmetry for the $a$'th kink, 
\be
k_{F_a}=\frac{4}{\pi\lp^3} \lambda(n_a)\, ,
\ee
in agreement with the result we used from \cite{Gaiotto:2009gz}.

%%%%%%%%%%%%%%%%%%%

%%%%%%%%%%%%%%%%%%%%%%%%%%%%%%%%%%%%%%%%%
%%%%%%%%%%%%%%%%%%%%%%%%%%%%%%%%%%%%%%%%%

\bibliographystyle{JHEP}

\bibliography{H2spindles}

\providecommand{\href}[2]{#2}\begingroup\raggedright\begin{thebibliography}{10}

\bibitem{Maldacena:2000mw}
J.~M. Maldacena and C.~Nunez, \emph{{Supergravity description of field theories
  on curved manifolds and a no go theorem}},
  \href{http://dx.doi.org/10.1142/S0217751X01003937}{\emph{Int. J. Mod. Phys.
  A} {\bf 16} (2001) 822--855},
  [\href{https://arxiv.org/abs/hep-th/0007018}{{\tt hep-th/0007018}}].

\bibitem{Gaiotto:2009gz}
D.~Gaiotto and J.~Maldacena, \emph{{The Gravity duals of N=2 superconformal
  field theories}},
  \href{http://dx.doi.org/10.1007/JHEP10(2012)189}{\emph{JHEP} {\bf 10} (2012)
  189}, [\href{https://arxiv.org/abs/0904.4466}{{\tt 0904.4466}}].

\bibitem{Gaiotto:2009we}
D.~Gaiotto, \emph{{N=2 dualities}},
  \href{http://dx.doi.org/10.1007/JHEP08(2012)034}{\emph{JHEP} {\bf 08} (2012)
  034}, [\href{https://arxiv.org/abs/0904.2715}{{\tt 0904.2715}}].

\bibitem{Reid-Edwards:2010vpm}
R.~A. Reid-Edwards and B.~Stefanski, jr., \emph{{On Type IIA geometries dual to
  N = 2 SCFTs}},
  \href{http://dx.doi.org/10.1016/j.nuclphysb.2011.04.002}{\emph{Nucl. Phys. B}
  {\bf 849} (2011) 549--572}, [\href{https://arxiv.org/abs/1011.0216}{{\tt
  1011.0216}}].

\bibitem{Aharony:2012tz}
O.~Aharony, L.~Berdichevsky and M.~Berkooz, \emph{{4d N=2 superconformal linear
  quivers with type IIA duals}},
  \href{http://dx.doi.org/10.1007/JHEP08(2012)131}{\emph{JHEP} {\bf 08} (2012)
  131}, [\href{https://arxiv.org/abs/1206.5916}{{\tt 1206.5916}}].

\bibitem{Gauntlett:2004zh}
J.~P. Gauntlett, D.~Martelli, J.~Sparks and D.~Waldram, \emph{{Supersymmetric
  AdS(5) solutions of M theory}},
  \href{http://dx.doi.org/10.1088/0264-9381/21/18/005}{\emph{Class. Quant.
  Grav.} {\bf 21} (2004) 4335--4366},
  [\href{https://arxiv.org/abs/hep-th/0402153}{{\tt hep-th/0402153}}].

\bibitem{Agarwal:2014rua}
P.~Agarwal, I.~Bah, K.~Maruyoshi and J.~Song, \emph{{Quiver tails and $
  \mathcal{N}=1 $ SCFTs from M5-branes}},
  \href{http://dx.doi.org/10.1007/JHEP03(2015)049}{\emph{JHEP} {\bf 03} (2015)
  049}, [\href{https://arxiv.org/abs/1409.1908}{{\tt 1409.1908}}].

\bibitem{Bah:2013aha}
I.~Bah and N.~Bobev, \emph{{Linear quivers and $ \mathcal{N} $ = 1 SCFTs from
  M5-branes}}, \href{http://dx.doi.org/10.1007/JHEP08(2014)121}{\emph{JHEP}
  {\bf 08} (2014) 121}, [\href{https://arxiv.org/abs/1307.7104}{{\tt
  1307.7104}}].

\bibitem{Bah:2012dg}
I.~Bah, C.~Beem, N.~Bobev and B.~Wecht, \emph{{Four-Dimensional SCFTs from
  M5-Branes}}, \href{http://dx.doi.org/10.1007/JHEP06(2012)005}{\emph{JHEP}
  {\bf 06} (2012) 005}, [\href{https://arxiv.org/abs/1203.0303}{{\tt
  1203.0303}}].

\bibitem{Anderson:2011cz}
M.~T. Anderson, C.~Beem, N.~Bobev and L.~Rastelli, \emph{{Holographic
  Uniformization}},
  \href{http://dx.doi.org/10.1007/s00220-013-1675-4}{\emph{Commun. Math. Phys.}
  {\bf 318} (2013) 429--471}, [\href{https://arxiv.org/abs/1109.3724}{{\tt
  1109.3724}}].

\bibitem{Bobev:2020jlb}
N.~Bobev, F.~F. Gautason and K.~Parmentier, \emph{{Holographic Uniformization
  and Black Hole Attractors}},
  \href{http://dx.doi.org/10.1007/JHEP06(2020)095}{\emph{JHEP} {\bf 06} (2020)
  095}, [\href{https://arxiv.org/abs/2004.05110}{{\tt 2004.05110}}].

\bibitem{Ferrero:2020laf}
P.~Ferrero, J.~P. Gauntlett, J.~M. P\'erez Ipi\~na, D.~Martelli and J.~Sparks,
  \emph{{D3-Branes Wrapped on a Spindle}},
  \href{http://dx.doi.org/10.1103/PhysRevLett.126.111601}{\emph{Phys. Rev.
  Lett.} {\bf 126} (2021) 111601},
  [\href{https://arxiv.org/abs/2011.10579}{{\tt 2011.10579}}].

\bibitem{Hosseini:2021fge}
S.~M. Hosseini, K.~Hristov and A.~Zaffaroni, \emph{{Rotating multi-charge
  spindles and their microstates}},
  \href{http://dx.doi.org/10.1007/JHEP07(2021)182}{\emph{JHEP} {\bf 07} (2021)
  182}, [\href{https://arxiv.org/abs/2104.11249}{{\tt 2104.11249}}].

\bibitem{Boido:2021szx}
A.~Boido, J.~M.~P. Ipi\~na and J.~Sparks, \emph{{Twisted D3-brane and M5-brane
  compactifications from multi-charge spindles}},
  \href{http://dx.doi.org/10.1007/JHEP07(2021)222}{\emph{JHEP} {\bf 07} (2021)
  222}, [\href{https://arxiv.org/abs/2104.13287}{{\tt 2104.13287}}].

\bibitem{Faedo:2021kur}
F.~Faedo, S.~Klemm and A.~Vigan\`o, \emph{{Supersymmetric black holes with
  spiky horizons}},
  \href{http://dx.doi.org/10.1007/JHEP09(2021)102}{\emph{JHEP} {\bf 09} (2021)
  102}, [\href{https://arxiv.org/abs/2105.02902}{{\tt 2105.02902}}].

\bibitem{Cassani:2021dwa}
D.~Cassani, J.~P. Gauntlett, D.~Martelli and J.~Sparks, \emph{{Thermodynamics
  of accelerating and supersymmetric AdS4 black holes}},
  \href{http://dx.doi.org/10.1103/PhysRevD.104.086005}{\emph{Phys. Rev. D} {\bf
  104} (2021) 086005}, [\href{https://arxiv.org/abs/2106.05571}{{\tt
  2106.05571}}].

\bibitem{Ferrero:2021ovq}
P.~Ferrero, M.~Inglese, D.~Martelli and J.~Sparks, \emph{{Multi-charge
  accelerating black holes and spinning spindles}},
  \href{https://arxiv.org/abs/2109.14625}{{\tt 2109.14625}}.

\bibitem{Couzens:2021rlk}
C.~Couzens, K.~Stemerdink and D.~van~de Heisteeg, \emph{{M2-branes on Discs and
  Multi-Charged Spindles}},  \href{https://arxiv.org/abs/2110.00571}{{\tt
  2110.00571}}.

\bibitem{Faedo:2021nub}
F.~Faedo and D.~Martelli, \emph{{D4-branes wrapped on a spindle}},
  \href{http://dx.doi.org/10.1007/JHEP02(2022)101}{\emph{JHEP} {\bf 02} (2022)
  101}, [\href{https://arxiv.org/abs/2111.13660}{{\tt 2111.13660}}].

\bibitem{Ferrero:2021etw}
P.~Ferrero, J.~P. Gauntlett and J.~Sparks, \emph{{Supersymmetric spindles}},
  \href{http://dx.doi.org/10.1007/JHEP01(2022)102}{\emph{JHEP} {\bf 01} (2022)
  102}, [\href{https://arxiv.org/abs/2112.01543}{{\tt 2112.01543}}].

\bibitem{Couzens:2021cpk}
C.~Couzens, \emph{{A tale of (M)2 twists}},
  \href{http://dx.doi.org/10.1007/JHEP03(2022)078}{\emph{JHEP} {\bf 03} (2022)
  078}, [\href{https://arxiv.org/abs/2112.04462}{{\tt 2112.04462}}].

\bibitem{Giri:2021xta}
S.~Giri, \emph{{Black holes with spindles at the horizon}},
  \href{https://arxiv.org/abs/2112.04431}{{\tt 2112.04431}}.

\bibitem{Bah:2021mzw}
I.~Bah, F.~Bonetti, R.~Minasian and E.~Nardoni, \emph{{Holographic Duals of
  Argyres-Douglas Theories}},
  \href{http://dx.doi.org/10.1103/PhysRevLett.127.211601}{\emph{Phys. Rev.
  Lett.} {\bf 127} (2021) 211601},
  [\href{https://arxiv.org/abs/2105.11567}{{\tt 2105.11567}}].

\bibitem{Bah:2021hei}
I.~Bah, F.~Bonetti, R.~Minasian and E.~Nardoni, \emph{{M5-brane sources,
  holography, and Argyres-Douglas theories}},
  \href{http://dx.doi.org/10.1007/JHEP11(2021)140}{\emph{JHEP} {\bf 11} (2021)
  140}, [\href{https://arxiv.org/abs/2106.01322}{{\tt 2106.01322}}].

\bibitem{Couzens:2021tnv}
C.~Couzens, N.~T. Macpherson and A.~Passias, \emph{{$ \mathcal{N} $ = (2, 2)
  AdS$_{3}$ from D3-branes wrapped on Riemann surfaces}},
  \href{http://dx.doi.org/10.1007/JHEP02(2022)189}{\emph{JHEP} {\bf 02} (2022)
  189}, [\href{https://arxiv.org/abs/2107.13562}{{\tt 2107.13562}}].

\bibitem{Suh:2021ifj}
M.~Suh, \emph{{D3-branes and M5-branes wrapped on a topological disc}},
  \href{http://dx.doi.org/10.1007/JHEP03(2022)043}{\emph{JHEP} {\bf 03} (2022)
  043}, [\href{https://arxiv.org/abs/2108.01105}{{\tt 2108.01105}}].

\bibitem{Suh:2021hef}
M.~Suh, \emph{{M2-branes wrapped on a topological disc}},
  \href{https://arxiv.org/abs/2109.13278}{{\tt 2109.13278}}.

\bibitem{Suh:2021aik}
M.~Suh, \emph{{D4-D8-branes wrapped on a manifold with non-constant
  curvature}},  \href{https://arxiv.org/abs/2108.08326}{{\tt 2108.08326}}.

\bibitem{Karndumri:2022wpu}
P.~Karndumri and P.~Nuchino, \emph{{Five-branes wrapped on topological disks
  from 7D N=2 gauged supergravity}},
  \href{http://dx.doi.org/10.1103/PhysRevD.105.066010}{\emph{Phys. Rev. D} {\bf
  105} (2022) 066010}, [\href{https://arxiv.org/abs/2201.05037}{{\tt
  2201.05037}}].

\bibitem{Anabalon:2018qfv}
A.~Anabal\'on, F.~Gray, R.~Gregory, D.~Kubiz\v{n}\'ak and R.~B. Mann,
  \emph{{Thermodynamics of Charged, Rotating, and Accelerating Black Holes}},
  \href{http://dx.doi.org/10.1007/JHEP04(2019)096}{\emph{JHEP} {\bf 04} (2019)
  096}, [\href{https://arxiv.org/abs/1811.04936}{{\tt 1811.04936}}].

\bibitem{Anabalon:2018ydc}
A.~Anabal\'on, M.~Appels, R.~Gregory, D.~Kubiz\v{n}\'ak, R.~B. Mann and
  A.~Ovg\"un, \emph{{Holographic Thermodynamics of Accelerating Black Holes}},
  \href{http://dx.doi.org/10.1103/PhysRevD.98.104038}{\emph{Phys. Rev. D} {\bf
  98} (2018) 104038}, [\href{https://arxiv.org/abs/1805.02687}{{\tt
  1805.02687}}].

\bibitem{Gregory:2017ogk}
R.~Gregory, \emph{{Accelerating Black Holes}},
  \href{http://dx.doi.org/10.1088/1742-6596/942/1/012002}{\emph{J. Phys. Conf.
  Ser.} {\bf 942} (2017) 012002}, [\href{https://arxiv.org/abs/1712.04992}{{\tt
  1712.04992}}].

\bibitem{Cheung:2022ilc}
K.~C.~M. Cheung, J.~H.~T. Fry, J.~P. Gauntlett and J.~Sparks, \emph{{M5-branes
  wrapped on four-dimensional orbifolds}},
  \href{https://arxiv.org/abs/2204.02990}{{\tt 2204.02990}}.

\bibitem{Couzens:2022lvg}
C.~Couzens, H.~Kim, N.~Kim, Y.~Lee and M.~Suh, \emph{{D4-branes wrapped on
  four-dimensional orbifolds through consistent truncation}},
  \href{https://arxiv.org/abs/2210.15695}{{\tt 2210.15695}}.

\bibitem{Lin:2004nb}
H.~Lin, O.~Lunin and J.~M. Maldacena, \emph{{Bubbling AdS space and 1/2 BPS
  geometries}},
  \href{http://dx.doi.org/10.1088/1126-6708/2004/10/025}{\emph{JHEP} {\bf 10}
  (2004) 025}, [\href{https://arxiv.org/abs/hep-th/0409174}{{\tt
  hep-th/0409174}}].

\bibitem{Ferrero:2021wvk}
P.~Ferrero, J.~P. Gauntlett, D.~Martelli and J.~Sparks, \emph{{M5-branes
  wrapped on a spindle}},
  \href{http://dx.doi.org/10.1007/JHEP11(2021)002}{\emph{JHEP} {\bf 11} (2021)
  002}, [\href{https://arxiv.org/abs/2105.13344}{{\tt 2105.13344}}].

\bibitem{Cvetic:1999xp}
M.~Cvetic, M.~J. Duff, P.~Hoxha, J.~T. Liu, H.~Lu, J.~X. Lu et~al.,
  \emph{{Embedding AdS black holes in ten-dimensions and eleven-dimensions}},
  \href{http://dx.doi.org/10.1016/S0550-3213(99)00419-8}{\emph{Nucl. Phys. B}
  {\bf 558} (1999) 96--126}, [\href{https://arxiv.org/abs/hep-th/9903214}{{\tt
  hep-th/9903214}}].

\bibitem{Lozano:2016kum}
Y.~Lozano and C.~N\'u\~nez, \emph{{Field theory aspects of non-Abelian
  T-duality and $ \mathcal{N} =$ 2 linear quivers}},
  \href{http://dx.doi.org/10.1007/JHEP05(2016)107}{\emph{JHEP} {\bf 05} (2016)
  107}, [\href{https://arxiv.org/abs/1603.04440}{{\tt 1603.04440}}].

\bibitem{Gauntlett:2006ai}
J.~P. Gauntlett, E.~O~Colgain and O.~Varela, \emph{{Properties of some
  conformal field theories with M-theory duals}},
  \href{http://dx.doi.org/10.1088/1126-6708/2007/02/049}{\emph{JHEP} {\bf 02}
  (2007) 049}, [\href{https://arxiv.org/abs/hep-th/0611219}{{\tt
  hep-th/0611219}}].

\bibitem{Nunez:2019gbg}
C.~N\'u\~nez, D.~Roychowdhury, S.~Speziali and S.~Zacar\'\i{}as,
  \emph{{Holographic aspects of four dimensional ${\cal N }=2$ SCFTs and their
  marginal deformations}},
  \href{http://dx.doi.org/10.1016/j.nuclphysb.2019.114617}{\emph{Nucl. Phys. B}
  {\bf 943} (2019) 114617}, [\href{https://arxiv.org/abs/1901.02888}{{\tt
  1901.02888}}].

\bibitem{Argyres:1995jj}
P.~C. Argyres and M.~R. Douglas, \emph{{New phenomena in SU(3) supersymmetric
  gauge theory}},
  \href{http://dx.doi.org/10.1016/0550-3213(95)00281-V}{\emph{Nucl. Phys. B}
  {\bf 448} (1995) 93--126}, [\href{https://arxiv.org/abs/hep-th/9505062}{{\tt
  hep-th/9505062}}].

\bibitem{Argyres:1995xn}
P.~C. Argyres, M.~R. Plesser, N.~Seiberg and E.~Witten, \emph{{New N=2
  superconformal field theories in four-dimensions}},
  \href{http://dx.doi.org/10.1016/0550-3213(95)00671-0}{\emph{Nucl. Phys. B}
  {\bf 461} (1996) 71--84}, [\href{https://arxiv.org/abs/hep-th/9511154}{{\tt
  hep-th/9511154}}].

\bibitem{Eguchi:1996vu}
T.~Eguchi, K.~Hori, K.~Ito and S.-K. Yang, \emph{{Study of N=2 superconformal
  field theories in four-dimensions}},
  \href{http://dx.doi.org/10.1016/0550-3213(96)00188-5}{\emph{Nucl. Phys. B}
  {\bf 471} (1996) 430--444}, [\href{https://arxiv.org/abs/hep-th/9603002}{{\tt
  hep-th/9603002}}].

\bibitem{Shapere:1999xr}
A.~D. Shapere and C.~Vafa, \emph{{BPS structure of Argyres-Douglas
  superconformal theories}},  \href{https://arxiv.org/abs/hep-th/9910182}{{\tt
  hep-th/9910182}}.

\bibitem{Cecotti:2010fi}
S.~Cecotti, A.~Neitzke and C.~Vafa, \emph{{R-Twisting and 4d/2d
  Correspondences}},  \href{https://arxiv.org/abs/1006.3435}{{\tt 1006.3435}}.

\bibitem{Xie:2012hs}
D.~Xie, \emph{{General Argyres-Douglas Theory}},
  \href{http://dx.doi.org/10.1007/JHEP01(2013)100}{\emph{JHEP} {\bf 01} (2013)
  100}, [\href{https://arxiv.org/abs/1204.2270}{{\tt 1204.2270}}].

\bibitem{Cecotti:2012jx}
S.~Cecotti and M.~Del~Zotto, \emph{{Infinitely many N=2 SCFT with ADE flavor
  symmetry}}, \href{http://dx.doi.org/10.1007/JHEP01(2013)191}{\emph{JHEP} {\bf
  01} (2013) 191}, [\href{https://arxiv.org/abs/1210.2886}{{\tt 1210.2886}}].

\bibitem{Wang:2015mra}
Y.~Wang and D.~Xie, \emph{{Classification of Argyres-Douglas theories from M5
  branes}}, \href{http://dx.doi.org/10.1103/PhysRevD.94.065012}{\emph{Phys.
  Rev. D} {\bf 94} (2016) 065012},
  [\href{https://arxiv.org/abs/1509.00847}{{\tt 1509.00847}}].

\bibitem{Maruyoshi:2016tqk}
K.~Maruyoshi and J.~Song, \emph{{Enhancement of Supersymmetry via
  Renormalization Group Flow and the Superconformal Index}},
  \href{http://dx.doi.org/10.1103/PhysRevLett.118.151602}{\emph{Phys. Rev.
  Lett.} {\bf 118} (2017) 151602},
  [\href{https://arxiv.org/abs/1606.05632}{{\tt 1606.05632}}].

\bibitem{Maruyoshi:2016aim}
K.~Maruyoshi and J.~Song, \emph{{$ \mathcal{N}=1 $ deformations and RG flows of
  $ \mathcal{N}=2 $ SCFTs}},
  \href{http://dx.doi.org/10.1007/JHEP02(2017)075}{\emph{JHEP} {\bf 02} (2017)
  075}, [\href{https://arxiv.org/abs/1607.04281}{{\tt 1607.04281}}].

\bibitem{Agarwal:2016pjo}
P.~Agarwal, K.~Maruyoshi and J.~Song, \emph{{$ \mathcal{N} $ =1 Deformations
  and RG flows of $ \mathcal{N} $ =2 SCFTs, part II: non-principal
  deformations}}, \href{http://dx.doi.org/10.1007/JHEP12(2016)103}{\emph{JHEP}
  {\bf 12} (2016) 103}, [\href{https://arxiv.org/abs/1610.05311}{{\tt
  1610.05311}}].

\bibitem{Agarwal:2017roi}
P.~Agarwal, A.~Sciarappa and J.~Song, \emph{{$ \mathcal{N} $ =1 Lagrangians for
  generalized Argyres-Douglas theories}},
  \href{http://dx.doi.org/10.1007/JHEP10(2017)211}{\emph{JHEP} {\bf 10} (2017)
  211}, [\href{https://arxiv.org/abs/1707.04751}{{\tt 1707.04751}}].

\bibitem{Benvenuti:2017bpg}
S.~Benvenuti and S.~Giacomelli, \emph{{Lagrangians for generalized
  Argyres-Douglas theories}},
  \href{http://dx.doi.org/10.1007/JHEP10(2017)106}{\emph{JHEP} {\bf 10} (2017)
  106}, [\href{https://arxiv.org/abs/1707.05113}{{\tt 1707.05113}}].

\bibitem{Bonelli:2011aa}
G.~Bonelli, K.~Maruyoshi and A.~Tanzini, \emph{{Wild Quiver Gauge Theories}},
  \href{http://dx.doi.org/10.1007/JHEP02(2012)031}{\emph{JHEP} {\bf 02} (2012)
  031}, [\href{https://arxiv.org/abs/1112.1691}{{\tt 1112.1691}}].

\bibitem{Giacomelli:2020ryy}
S.~Giacomelli, N.~Mekareeya and M.~Sacchi, \emph{{New aspects of
  Argyres--Douglas theories and their dimensional reduction}},
  \href{http://dx.doi.org/10.1007/JHEP03(2021)242}{\emph{JHEP} {\bf 03} (2021)
  242}, [\href{https://arxiv.org/abs/2012.12852}{{\tt 2012.12852}}].

\bibitem{Bah:2018gwc}
I.~Bah and E.~Nardoni, \emph{{Structure of Anomalies of 4d SCFTs from
  M5-branes, and Anomaly Inflow}},
  \href{http://dx.doi.org/10.1007/JHEP03(2019)024}{\emph{JHEP} {\bf 03} (2019)
  024}, [\href{https://arxiv.org/abs/1803.00136}{{\tt 1803.00136}}].

\bibitem{Xie:2013jc}
D.~Xie and P.~Zhao, \emph{{Central charges and RG flow of strongly-coupled N=2
  theory}}, \href{http://dx.doi.org/10.1007/JHEP03(2013)006}{\emph{JHEP} {\bf
  03} (2013) 006}, [\href{https://arxiv.org/abs/1301.0210}{{\tt 1301.0210}}].

\bibitem{Kang:2021lic}
M.~J. Kang, C.~Lawrie and J.~Song, \emph{{Infinitely many 4D N=2 SCFTs with a=c
  and beyond}},
  \href{http://dx.doi.org/10.1103/PhysRevD.104.105005}{\emph{Phys. Rev. D} {\bf
  104} (2021) 105005}, [\href{https://arxiv.org/abs/2106.12579}{{\tt
  2106.12579}}].

\bibitem{Nishinaka:2012vi}
T.~Nishinaka, \emph{{The gravity duals of SO/USp superconformal quivers}},
  \href{http://dx.doi.org/10.1007/JHEP07(2012)080}{\emph{JHEP} {\bf 07} (2012)
  080}, [\href{https://arxiv.org/abs/1202.6613}{{\tt 1202.6613}}].

\bibitem{Freed:1998tg}
D.~Freed, J.~A. Harvey, R.~Minasian and G.~W. Moore, \emph{{Gravitational
  anomaly cancellation for M theory five-branes}},
  \href{http://dx.doi.org/10.4310/ATMP.1998.v2.n3.a8}{\emph{Adv. Theor. Math.
  Phys.} {\bf 2} (1998) 601--618},
  [\href{https://arxiv.org/abs/hep-th/9803205}{{\tt hep-th/9803205}}].

\bibitem{Harvey:1998bx}
J.~A. Harvey, R.~Minasian and G.~W. Moore, \emph{{NonAbelian tensor multiplet
  anomalies}},
  \href{http://dx.doi.org/10.1088/1126-6708/1998/09/004}{\emph{JHEP} {\bf 09}
  (1998) 004}, [\href{https://arxiv.org/abs/hep-th/9808060}{{\tt
  hep-th/9808060}}].

\bibitem{Bah:2019rgq}
I.~Bah, F.~Bonetti, R.~Minasian and E.~Nardoni, \emph{{Anomalies of QFTs from
  M-theory and Holography}},
  \href{http://dx.doi.org/10.1007/JHEP01(2020)125}{\emph{JHEP} {\bf 01} (2020)
  125}, [\href{https://arxiv.org/abs/1910.04166}{{\tt 1910.04166}}].

\bibitem{10.4310/jdg/1214460608}
R.~Bott and A.~S. Cattaneo, \emph{{Integral invariants of {3}-manifolds}},
  \href{http://dx.doi.org/10.4310/jdg/1214460608}{\emph{J. Diff. Geom.} {\bf
  48} (1998) 91 -- 133}, [\href{https://arxiv.org/abs/dg-ga/9710001}{{\tt
  dg-ga/9710001}}].

\end{thebibliography}\endgroup
%%%%%%%%%%%%%%%%%%%%%%%%%%%%%%%%%%%%%%%%
%%%%%%%%%%%%%%%%%%%%%%%%%%%%%%%%%%%%%%%%%

%%%%%%%%%%%%%%%%%%%%%%%%%%%%%%%%%%%%
%%%%%%%%%%%%%%%%%%%%%%%%%%%%%%%%%%%%
\end{document}